\documentclass[aps,prd,twocolumn,showpacs,superscriptaddress,
 groupedaddress,nofootinbib,longbibliography]{revtex4-1}  
\usepackage[dvips]{graphicx}  

\usepackage{bm}        
\usepackage{amssymb}   
\usepackage{amsmath}   

\usepackage{tikz}
\usetikzlibrary{shapes,arrows}
\usetikzlibrary{positioning,decorations.text,decorations.pathmorphing}
\usetikzlibrary{spy}

\tikzstyle{block} = [rectangle, draw, fill=blue!10, 
    text width=5em, text centered, rounded corners, minimum height=2.5ex]
\tikzstyle{wblock} = [rectangle, draw, fill=blue!10, 
    text width=8em, text centered, rounded corners, minimum height=2.5ex]
\tikzstyle{sblock} = [rectangle, draw, fill=blue!10, 
    text width=3em, text centered, rounded corners, minimum height=2.5ex]
\tikzstyle{line} = [draw, -latex']

\begin{document}

\preprint{TUM-EFT 76/15}

\title{Polyakov loop in 2+1 flavor QCD from low to high temperatures}

\author{A. Bazavov$^a$, N. Brambilla$^{b,c}$, H.-T. Ding$^d$, 
P. Petreczky$^e$, H.-P. Schadler$^{e,f}$, A. Vairo$^b$,
J. H. Weber$^{b,g}$ \\ (TUMQCD Collaboration)}
\affiliation{
$^a$ Department of Physics and Astronomy, University of Iowa, 
Iowa City, Iowa 52242-1479, USA\\
$^b$ Physik Department, Technische Universit\"{a}t M\"{u}nchen, 
D-85748 Garching, Germany\\
$^c$ Institute of Advanced Studies, Technische Universit\"{a}t M\"{u}nchen, 
D-85748 Garching, Germany\\
$^d$ Key Laboratory of Quark \& Lepton Physics (MOE) and Institute of 
Particle Physics, Central China Normal University, Wuhan 430079, China\\
$^e$ Physics Department, Brookhaven National Laboratory, Upton, New York 11973, USA\\
$^f$ Institute of Physics, University of Graz, 8010 Graz, Austria\\
$^g$ Exzellenzcluster Universe, Technische Universit\"{a}t M\"{u}nchen, 
D-85748 Garching, Germany
}
\date{\today}

\begin{abstract}
We study the free energy of a static quark in QCD with 2+1 flavors in a 
wide temperature region, 116 MeV $< T < $ 5814 MeV, using the highly improved 
staggered quark (HISQ) action. 
We analyze the transition region in detail, obtain the entropy of a static 
quark, show that it peaks at temperatures close to the chiral crossover 
temperature and also 
revisit the  temperature dependence of the Polyakov loop susceptibilities 
using gradient flow.
We discuss the implications of our findings for the deconfinement and 
chiral crossover phenomena at physical values of the quark masses. 
Finally a  comparison of the lattice results at high temperatures with 
the weak-coupling calculations is presented.
\end{abstract}

\pacs{12.38.Gc, 12.38.-t, 12.38.Bx, 12.38.Mh}
\maketitle

\section{\label{sec:Introduction} Introduction}
{
As the temperature is increased, strongly interacting matter undergoes 
a transition to a state with different properties than the vacuum at zero 
temperature. 
Deconfinement of gluons and quarks, restoration of chiral symmetry and 
screening of color charges are the key properties of this thermal medium 
(for recent reviews see \mbox{e.g.}~\cite{Bazavov:2015rfa, Ding:2015ona, 
Petreczky:2012rq}).

The expectation value of the Polyakov loop is a sensitive probe of the 
screening properties of the medium. 
In SU(N) gauge theories the Polyakov loop is an order parameter for 
deconfinement.
At the transition temperature, both the bare and the renormalized Polyakov 
loop exhibit a discontinuity and their fluctuations diverge. 
Hence, the bare Polyakov loop is used to study the deconfinement phase 
transition in SU(N) gauge theories, in particular the bare Polyakov loop 
susceptibility is used to define the phase transition temperature (see e.g. 
Ref. \cite{Boyd:1996bx}). 
To what extent it is a sensitive probe of deconfinement in QCD with light 
dynamical quarks is not quite clear in view of the crossover nature of the 
transition \cite{Aoki:2006we}.  
In particular, it is not clear if it is possible to define a crossover 
temperature from the bare Polyakov loop, since it is a continuous quantity 
in the crossover region. 
In recent years the deconfinement transition in QCD with light dynamical 
quarks has been studied in terms of fluctuations and correlations of 
conserved charges, which indicate the appearance of quark degrees of 
freedom just above the chiral transition temperature \cite{Bazavov:2013dta, 
Bazavov:2014yba, Bellwied:2013cta, Mukherjee:2015mxc}.

After proper renormalization the expectation value of the renormalized 
Polyakov loop is related to the free energy, $F_Q$, of a static quark 
\cite{McLerran:1981pb,Kaczmarek:2002mc}
\begin{equation}
 L^{\rm{ren}}=\exp(-F_Q^{\rm{ren}}/T).  
\end{equation} 
The renormalized Polyakov loop, or equivalently the free energy of a 
static charge $F_Q$ has been studied in SU(N) gauge theories in a wide 
temperature interval \cite{Kaczmarek:2002mc, Digal:2003jc, Kaczmarek:2004gv, 
Gupta:2007ax, Mykkanen:2012ri}. 
Comparisons of the lattice results with weak-coupling calculations have 
also been performed up to next-to-leading order (NLO)~\cite{Burnier:2009bk} 
and up to next-to-next-to-leading order (NNLO)~\cite{Berwein:2015ayt}.

The renormalized Polyakov loop has been computed in QCD with dynamical 
quarks for various quark flavor content and quark masses~\cite{Petreczky:2004pz, 
Kaczmarek:2005ui, Cheng:2007jq, Bazavov:2009zn, Cheng:2009zi, Borsanyi:2010bp, 
Bazavov:2011nk, Bazavov:2013yv, Borsanyi:2015yka}. 
Continuum extrapolated results with physical quark masses exist for staggered 
fermion formulations \cite{Borsanyi:2010bp, Bazavov:2013yv, Borsanyi:2015yka}. 
For large quark masses continuum results are also available for overlap and 
Wilson fermion formulations \cite{Borsanyi:2015waa, Borsanyi:2015zva}. 
Unfortunately, none of the above studies extend to sufficiently high 
temperature to make contact with weak-coupling calculations. 

The relation of the Polyakov loop to the nature of the QCD crossover 
remains unclear. 
For large quark masses the deconfinement crossover defined in terms of 
the Polyakov loop and the chiral crossover defined in terms of the chiral 
condensate happen at about the same temperature \cite{Karsch:2000kv, 
Petreczky:2004pz, Kaczmarek:2005ui}. 
In the crossover region, both the Polyakov loop and the chiral condensate 
change rapidly and their fluctuations become large. 
For physical values of the quark masses the situation may be different. 
In Refs. \cite{Aoki:2006br, Aoki:2009sc} it was found that the deconfinement 
crossover defined in terms of the renormalized Polyakov loop happens at 
temperatures significantly higher than the chiral crossover temperature 
defined as the maximum of the chiral susceptibility. 
The study of ratios of fluctuations of the imaginary and real parts of the 
Polyakov loop in Ref. \cite{Lo:2013hla} suggested that the deconfinement and 
chiral crossover happen at about the same temperature. 
However, as this study used an ad-hoc renormalization prescription, lacked 
continuum extrapolation and provided no information on the cutoff effects in 
full QCD, the implications of this result are not conclusive. 

In this paper we will study the free energy of a static quark in a broad 
temperature region extending to $5.8$ GeV. 
We will also reexamine the behavior of $F_Q$ in the transition region, 
in particular,
we will calculate the entropy of a static quark, 
$S_Q=-\partial F_Q/\partial T$, and discuss its relation to the 
deconfinement transition temperature. 
We will show that the deconfinement transition temperature, defined at the 
peak of $S_Q$, is actually consistent with the chiral transition temperature. 

The rest of the paper is organized as follows. 
In Sec. II we discuss our lattice setup.
In Sec. III we discuss the renormalization of the Polyakov loop using the 
static quark antiquark energy at zero temperature. In Sec. IV our results 
on the entropy of a static quark will be presented. 
Sec. V will show how to extend the lattice calculations of
the static quark free energy to higher temperatures. 
In Sec. VI we will discuss the calculation of the renormalized Polyakov loop 
and its susceptibility using the gradient flow.
The free energy of a static quark in the high temperature region will be 
compared to the weak-coupling results in Sec. VII. 
Finally, Sec. VIII contains our conclusions. 
Some technical details of the calculations will be given in the appendices.
}

\section{\label{sec:Setup} Lattice QCD Setup}
{
\begin{figure}
 \includegraphics[width=8cm]{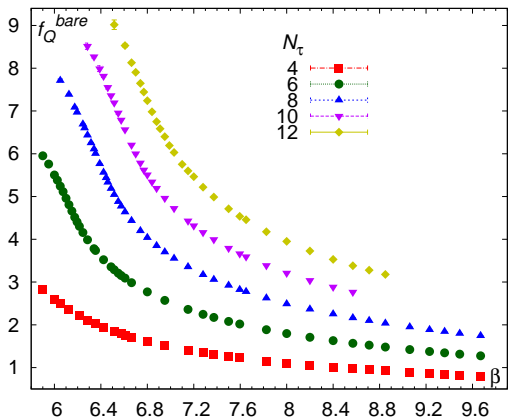}
  \caption{\label{fig: bare Fq} 
The bare free energy of a static quark  
$ f_Q^{\rm bare}=F_Q^{\rm{bare}}/T= -\log L^{\rm{bare}} $ as function of the 
gauge coupling $\beta$ for different $N_{\tau}$ values.}
\end{figure}
We perform calculations of the bare Polyakov loop at nonzero temperature
on $N_{\sigma}^3 \times N_{\tau}$ lattices with $N_\tau=4,~6,~8,~10$ and $12$, 
and the
aspect ratio of $N_{\sigma}/N_{\tau}=4$ using the highly improved staggered 
quark (HISQ) action~\cite{Follana:2006rc}. 
The gauge configurations have been generated by the HotQCD Collaboration
~\cite{Bazavov:2011nk,Bazavov:2014pvz}, in the course of studies of quark 
number susceptibilities at high temperatures \cite{Ding:2015fca, 
Bazavov:2013uja} as well as in
a previous study of the renormalized Polyakov loop with the HISQ action 
\cite{Bazavov:2013yv}. 

We required additional gauge configurations and generated these using the 
SuperMUC and C2PAP computers at Leibniz Rechenzentrum (LRZ) in Garching. 
Additional gauge configurations have been generated for $N_{\tau}=4,~6$ and 
$8$ to calculate the Polyakov loop at very high temperatures.
Further gauge configurations have been generated for $N_{\tau}=10$ and $12$ 
to reduce uncertainties of the free energy at low temperatures and achieve 
sufficient resolution of the peak of $ S_Q $. 

The gauge configurations have been generated in the range of gauge coupling 
$\beta=5.90-9.67$ with $ \beta=10/g_0^2$ using the rational hybrid 
Monte-Carlo (RHMC) algorithm and the MILC~code.
Details on the HISQ action implementation in the MILC~code can be found 
in~\cite{Bazavov:2010ru}. 
The lattice spacing $a$ has been fixed by the $r_1$ scale and we use the 
parametrization of $r_1/a$ given in Ref. \cite{Bazavov:2014pvz}. 
Using this parametrization we find that the above $\beta$ range 
corresponds to a temperature range of $116\ {\rm MeV} < T < 5814\ {\rm MeV}$. 
The Polyakov loop has been calculated after each molecular dynamic time 
unit (TU). 
For temperatures $T<407$ MeV the accumulated statistics corresponds to 
$30-60$ thousands of TUs. 
At higher temperatures in many cases far fewer gauge configurations are 
available. 
The details on collected statistics are given in Appendix A. 

The Polyakov loop on the lattice is defined as 
\begin{equation}
P({\bf x})=\frac{1}{3} {\rm Tr}\, 
\prod_{x_0=0}^{N_{\tau}-1} U_0({\bf x},x_0),
\label{defP}
\end{equation}
where $U_{\mu}(x=({\bf x},x_0))$ are the lattice link variables.
The bare expectation value of the Polyakov loop will be denoted by 
$L^{\rm{bare}}$ in what follows, $L^{\rm{bare}}= \langle P \rangle$. 
Since the expectation value of the Polyakov loop is independent of 
${\bf x}$ we average the Polyakov loop over the entire spatial volume.
Our results for the bare Polyakov loop are summarized in Fig. 
\ref{fig: bare Fq} in terms of the scaled bare static quark free energy  
$f_Q^{\mathrm{bare}}= -\log L^{\rm{bare}} $ 
as a function of the gauge coupling $ \beta $. 
Here and in what follows we denote by $f_Q^{\rm{bare}}$ the scaled bare 
free energy of a static quark,
$f_Q^{\rm{bare}}=F_Q^{\rm{bare}}/T$. 
As one may see from the figure, $f_Q^{\rm{bare}}$ decreases for increasing 
$\beta$ and for decreasing $N_{\tau}$. 
The continuum limit at fixed temperature would be reached by varying 
$N_{\tau}$ and $\beta$ simultaneously in the limit $N_{\tau} \to \infty$, 
following lines going from the lower left corner into the direction of the 
upper right corner. 
Since $f_Q^{\rm{bare}}$ diverges as one proceeds along these lines, the 
continuum limit of $f_Q^{\rm{bare}}$ is not defined. 
Thus, we must subtract this divergence before taking the continuum limit. 
We will discuss this in the next section. 
}

\section{\label{sec:renL}
Renormalization of the Polyakov loop and the continuum extrapolation}
{
The Polyakov loop needs multiplicative renormalization 
\cite{Polyakov:1980ca}.
This means that the free energy of a static quark $F_Q$ needs an additive 
renormalization.
The additive renormalization of $F_Q$ is related to the additive 
renormalization of the energy of a static quark antiquark ($Q \bar Q$) 
pair at zero temperature.
The static quark antiquark free energy $F_{Q\bar Q}(r,T)$ agrees with 
the static quark antiquark energy at zero temperature at short distances 
once a finite additive term due to trivial color factors is included 
\cite{Kaczmarek:2002mc}. 
On the other hand $F_{Q\bar Q}(r \rightarrow \infty,T)=2 F_Q(T)$ 
\cite{McLerran:1981pb, Kaczmarek:2002mc}. 
Therefore, the renormalization constant of $F_Q$, which we denote by $C_Q$, 
is half of the renormalization constant of the static energy at zero 
temperature.  

To determine the normalization constant $C_Q$ we require that the static 
$Q \bar Q$ energy for zero temperature at a distance $r=r_0$ is equal to 
$0.954/r_0$ \cite{Bazavov:2011nk}.
This normalization condition is equivalent to normalizing the static energy 
to $0.2065/r_1$ \cite{Bazavov:2014pvz} at a distance $r=r_1$.
Normalizing the static energy at $r_1$, \mbox{i.e.}~at shorter distances has 
the advantage of reducing the statistical errors at large $\beta$, while the 
normalization at distance $r_0$ is more suitable for coarser lattices, 
\mbox{i.e.}~smaller values of $\beta$.
Using the lattice results on the static $Q \bar Q$ energy from Ref. 
\cite{Bazavov:2011nk} and normalizing them to $0.954/r_0$ for 
$\beta \le 6.488$ we determine $r_0 C_Q$.
Then using the results on the static $Q \bar Q$ energy at higher $\beta$ 
from Refs. \cite{Bazavov:2011nk,Bazavov:2014pvz} and normalizing those 
to $0.2065/r_1$ we determine $r_1 C_Q$. 
Finally using $r_1/a$ and $r_0/a$ from Refs. \cite{Bazavov:2011nk, 
Bazavov:2014pvz} we calculate the values of the normalization constant 
in lattice units $a C_Q(\beta)=c_Q(\beta)$ which are shown in Fig. 
\ref{fig:stdzren} and tabulated in Appendix~\ref{appendix:lattice_details}.
Note that since $C_Q$ has a $1/a$ divergence, $c_Q$ is finite and is a 
slowly varying function of $\beta$. 
Once the cutoff dependence is rephrased in terms of the lattice spacing 
$a(\beta)$, we may write $C_Q = b/a + c + \mathcal{O}(a^2)$. 
The divergence $b/a$ cancels against the divergence of the bare free energy.
The constant $c$ is a scheme dependent constant, which depends on the 
distances $r_0$ or $r_1$, but is independent of the lattice spacing. 
Since the leading higher order corrections are suppressed by 
$\alpha_s a^2$ for the HISQ/Tree action, the derivative in $a$ of these 
corrections vanishes in the continuum limit. 
We note that, since $T=1/(aN_{\tau})$, at fixed $N_{\tau}$ the dependence 
of $c_Q$ on $a$ translates into a dependence on the temperature. 

\begin{figure}
  \includegraphics[width=8cm]{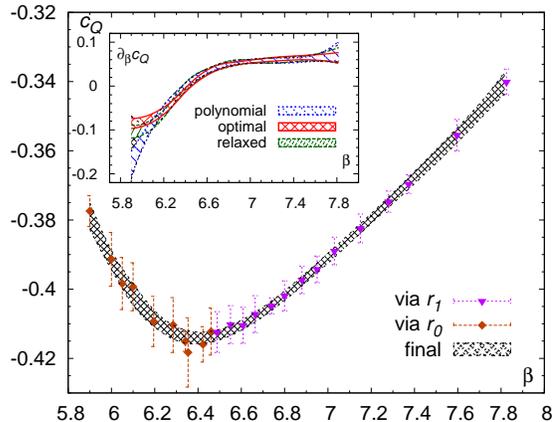}
  \caption{\label{fig:stdzren} 
  Renormalization constant $ c_Q(\beta) $ from the $ Q\bar Q $ 
  renormalization procedure. 
  Interpolations are shown as $ 1\sigma $ bands and data points are 
  explained in the text. 
  The inset shows the derivative $ \frac{\partial c_Q}{\partial \beta} $. 
  Optimal and relaxed refer to different spline interpolations with 
  $n_k=4$ or $n_k=5$ knots respectively. 
  }
\end{figure}
\begin{figure*}
\includegraphics[width=8cm]{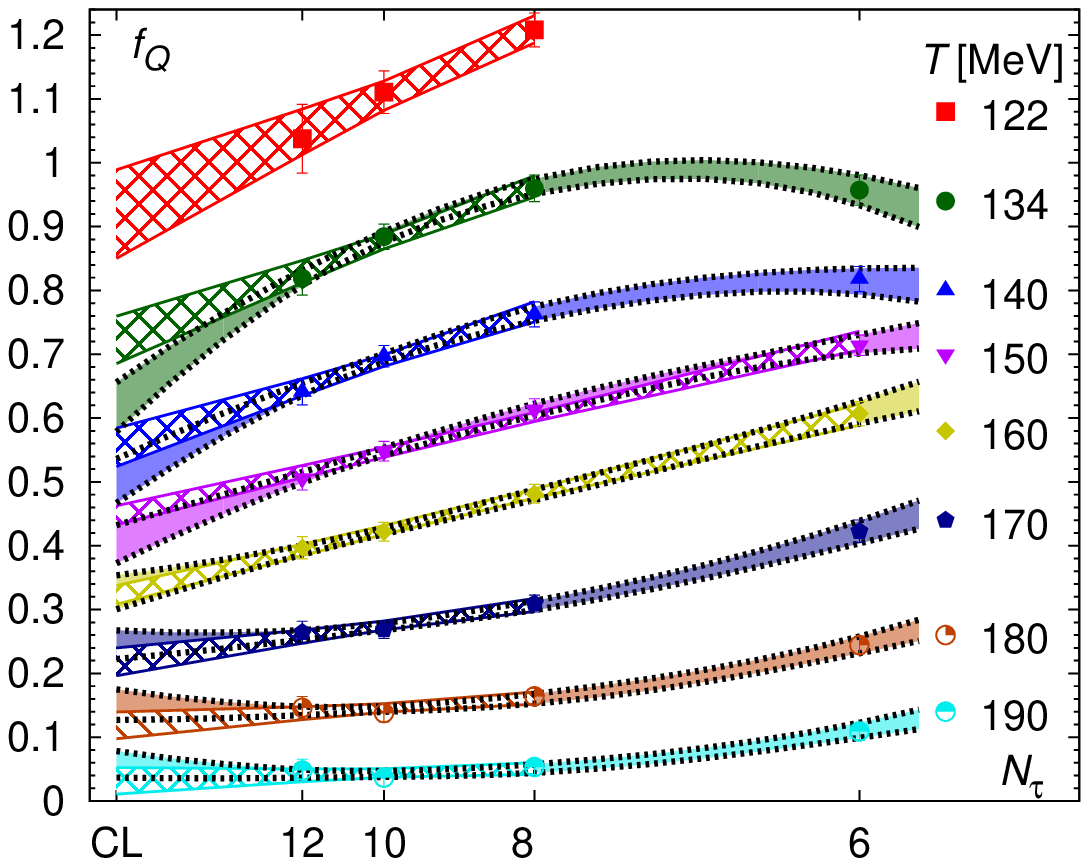}
\includegraphics[width=8cm]{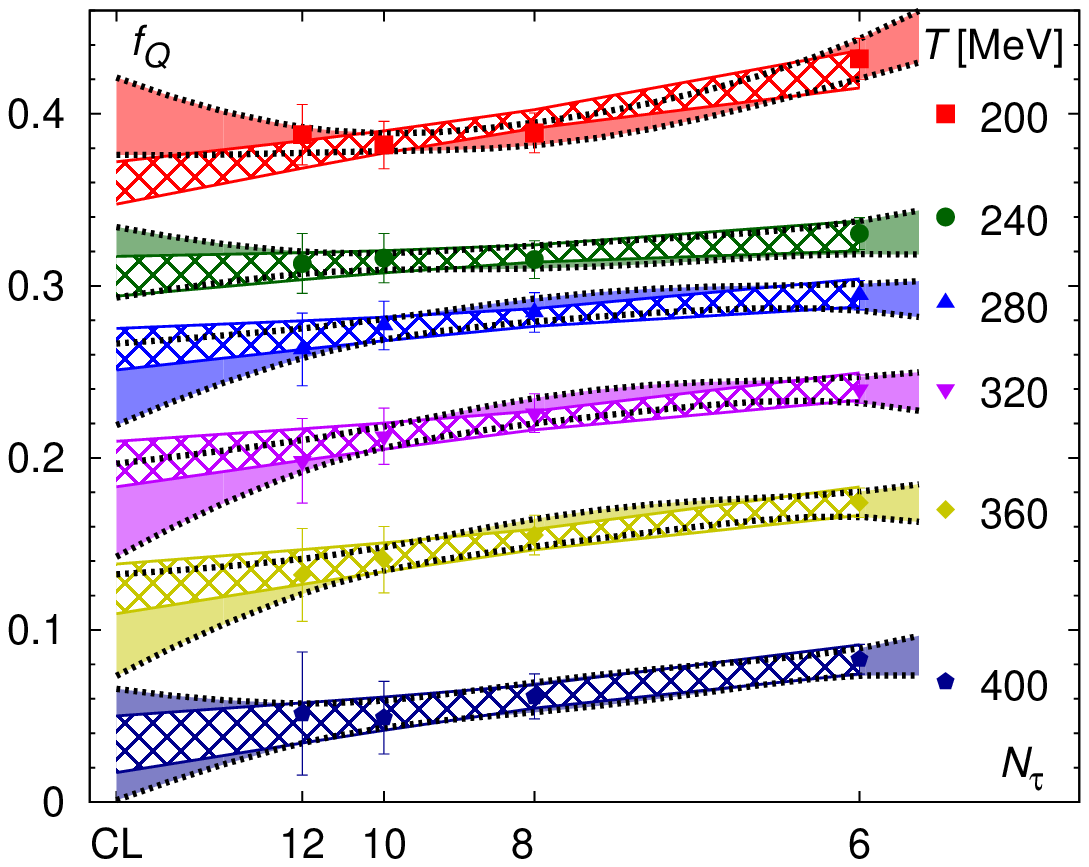}
\caption{ 
The static quark free energy at various temperatures as function of 
$N_{\tau}$. 
``CL'' marks the continuum limit ($N_\tau \to \infty$). 
Results for each temperature are shifted by some constant for better 
visibility. 
The $1/N_{\tau}^2$ continuum extrapolations are shown as bands with 
filled pattern. 
The continuum extrapolations with $1/N_{\tau}^4$ term included are shown 
as solid filled bands.
The width of the band shows the statistical uncertainty of the fits.
The left panel shows the results in the low temperature region, while 
the right panel shows the results in the high temperature region.
}
\label{fig:fNt}
\end{figure*}
\begin{figure}
\includegraphics[width=8cm]{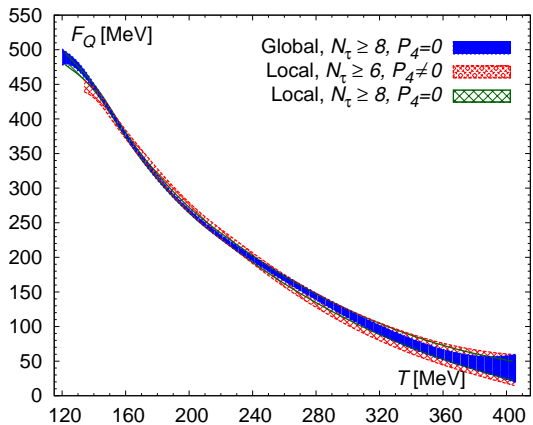}
\caption{ 
Different continuum extrapolations for the static quark free energy $F_Q$. 
We show extrapolations with coefficient $P_4/N_{\tau}^4$ term set to zero 
as well as
for nonzero values of the coefficient $P_4$. 
}
\label{fig:fcont}
\end{figure}
\begin{figure}
\includegraphics[width=8cm]{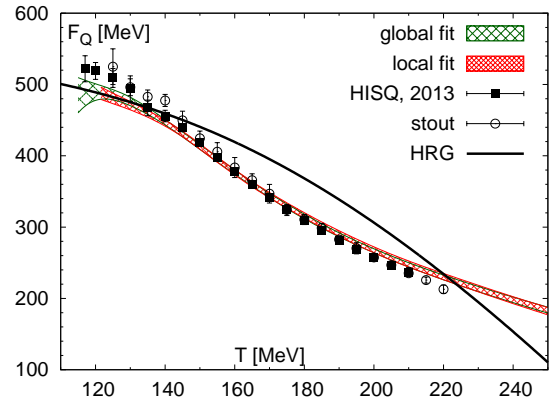}
\caption{ 
The continuum results for the free energy of static quark compared to 
previous calculations \cite{Borsanyi:2010bp,Bazavov:2013yv}. 
Also shown as a solid black line is the hadron resonance gas calculation of 
$F_Q$ from Ref. \cite{Bazavov:2013yv}.}
\label{fig:fcomp}
\end{figure}

Now for the renormalized free energy in temperature units we can write
\begin{equation}
  f_Q^{\mathrm{ren}}(T(\beta,N_{\tau}),N_{\tau}) 
  = f_Q^{\rm{bare}}(\beta,N_\tau) + N_\tau c_Q(\beta).
  \label{eq: fqren}
\end{equation}
The renormalized free energy depends on $\beta$ through the chain rule 
for $T(\beta,N_\tau)$. 
We use $T$ as argument instead of $\beta$, since the continuum limit of 
$f_Q^{\rm ren}(T(\beta,N_\tau),N_{\tau})$ can be taken for fixed temperature. 
Hereafter, we usually omit the superscript ``ren'' when referring to 
renormalized quantities, but keep the superscript ``bare'' for the bare 
quantities. 
Here and in what follows we denote by $f_Q$ the scaled renormalized 
free energy of a static quark,
$f_Q=F_Q/T$. 
In order to  determine $f_Q^{\rm{bare}}$ and $c_Q$ as a function of 
$\beta$ and/or as a function of the temperature, we interpolate the lattice 
results on $c_Q(\beta)$ and $f_Q^{\rm{bare}}(\beta,N_\tau)$ independently 
in $\beta$.

First, we discuss the interpolation procedure for $c_Q$.
To obtain $c_Q$ as a function of $\beta$ we use smooth splines and 
polynomial interpolations.
The errors on the interpolations have been estimated using the bootstrap 
method. 
We varied the number of knots of the splines as well as the value of the 
smoothing parameter in order to estimate the systematic errors. 
In the case of polynomial fits we consider polynomials of different degree. 
The interpolation of $c_Q$ is also shown in Fig. \ref{fig:stdzren}.
In the inset of the figure we show the derivative of $c_Q$ with respect to 
beta in order to highlight the spread in different interpolations.
The differences between the different interpolations are most visible in 
the $\beta$ dependence of the derivative of $c_Q$ that is needed for the e
valuation of the entropy of a static charge to be discussed in the next 
section. 

Next, we discuss the interpolations of the free energy as well as the 
continuum extrapolations. 
At finite cutoff, the temperature $T$ is related to $N_\tau$ and the lattice 
spacing $a$ through $a N_{\tau} =1/T$; 
trading $a$ for $\beta$ we can also write $\beta = \beta(T,N_\tau)$. 
Consequently, the limit $a \to 0$ at fixed temperature is tantamount to the 
limit $N_{\tau} \to \infty$. 
The power law dependence of cutoff effects on $a$ or $1/N_{\tau}$ respectively 
is determined by the leading discretization errors of the lattice simulations 
($\mathcal{O}(\alpha_s a^2,a^4)$ for the HISQ action). 
We will use two approaches to do this, which we will call local and global 
extrapolations. 
In the first approach, which we will call a local fit, we perform the 
interpolation of the lattice results for $f_Q^{\rm bare}$ as function of 
$\beta$ for each $N_{\tau}$ separately. 
Using the value of $c_Q$ determined above we then calculate the renormalized 
free energy $f_Q(T(\beta,N_\tau),N_{\tau})$ for each $N_{\tau}$ and perform 
continuum extrapolations. 
In the second approach, which we will call a global fit, we simultaneously 
fit the temperature and $N_{\tau}$ dependence of 
$f_Q^{\mathrm{bare}}(\beta,N_\tau) + N_\tau c_Q(\beta)$. 
Setting  $N_{\tau} \rightarrow \infty$ in the resulting fit we obtain 
the continuum extrapolated results for the renormalized free energy. 
We will discuss these two approaches in the following subsections in 
more details.

\subsection{\label{Local fits}Local interpolations and extrapolations}

To perform the interpolation of $f_Q^{\rm bare}(\beta,N_{\tau})$ we split 
the $\beta$ range in overlapping low $\beta$ and high $\beta$ intervals which 
roughly correspond to temperatures $T<200$ MeV and $T>200$ MeV respectively. 
In these intervals for each $N_{\tau}$ we perform interpolations in $\beta$ 
using smoothing splines as well as polynomial fits. 
We find that in the low beta range it is sufficient to use splines with $5-7$ 
knots, while in the high $\beta$ range we use splines with $8-19$  knots 
depending on the value of $N_{\tau}$. 
The statistical errors of the interpolations are estimated using the 
bootstrap method. 
To estimate possible systematic errors in the interpolation we also performed 
polynomial fits of the lattice data for $f_Q^{\rm bare}(\beta,N_{\tau})$ in 
the above intervals. 
We find that the interpolations obtained with polynomials and splines agree 
well within the estimated statistical errors not only for 
$f_Q^{\rm bare}(\beta,N_{\tau})$ but also for its derivative. 
Therefore, there are no additional systematic errors in our analysis.
The details of the interpolations and fits are presented in Appendix 
\ref{appendix:fits}.
Having the interpolation for $f_Q^{\rm bare}(\beta,N_{\tau})$ and the 
interpolation for $c_Q$ we  calculate the renormalized free energy for 
each $N_{\tau}$. 
We then perform a $1/N_{\tau}^2$ extrapolation for $f_Q$ to obtain the 
continuum limit for each value of the temperature. 
In Fig. \ref{fig:fNt} we show the $N_{\tau}$ dependence of $f_Q$ 
together with $1/N_{\tau}^2$ and $1/N_{\tau}^4$ extrapolations. 
As one can see from the figure cutoff effects are fairly small for 
$T>200$ MeV and $1/N_\tau^2$ holds including $N_{\tau}=6$ data. 
Note that we do not consider the $N_{\tau}=4$ results partly because 
they are available only for $T>200$ MeV and partly because they are 
outside the scaling window. 
At lower temperature cutoff effects are larger and the $N_{\tau}=6$ 
data are not in the scaling regime. 
Therefore, we have to consider fits with $1/N_{\tau}^4$ term included, 
or use $1/N_{\tau}^2$ fits for $N_{\tau} \geq 8$ only.
The continuum results obtained with the above extrapolations are shown 
in Fig. \ref{fig:fcont}.

\subsection{\label{Global fits}Global fits and extrapolations}

In the previous subsection we have seen that the temperature dependence 
can be described by polynomials in the low and high beta ranges once 
$\beta$ has been reexpressed in $T$. 
Furthermore, the $N_{\tau}$ dependence of the lattice results is well 
described by a function $P_0+P_2/N_{\tau}^2+P_4/N_{\tau}^4$. 
Therefore, we performed fits for $N_{\tau}=6,~8,~10$ and $12$ data 
on $f_Q(T(\beta,N_{\tau}),N_{\tau})$ 
using the following form
\begin{equation}
P_0(T)+\frac{P_2(T)}{N_{\tau}^2}+\frac{P_4(T)}{N_{\tau}^4}.
\label{eq:global fit}
\end{equation}
Here $P_i,~i=0,2,4$ are polynomials in the temperature $T$. 
As we did for local interpolations, we split the temperature range in 
overlapping low and high temperature intervals and performed the global 
fits in both intervals separately. 
These intervals roughly correspond to $T<200$ MeV and $T>200$ MeV. 
The low temperature fits extend only down to the lowest temperatures 
where bare free energies are available for $N_{\tau}=12$, which is 
slightly above $120$ MeV. 
The high temperature fits extend only up to the highest temperature 
where $c_Q$ is available for $N_{\tau}=12$, which is slightly below $410$ MeV.
We used fits with and without the $1/N_{\tau}^4$ term, as well as 
including
and excluding the $N_{\tau}=6$ data.  
We find that within estimated statistical errors all the fits agree 
both for $f_Q(T(\beta,N_\tau),N_{\tau})$ and its derivatives.
The account of these fits is given in appendix \ref{appendix:fits}.
For the continuum result we use the fit which does not include the 
$N_{\tau}=6$ data
and has fixed $P_4=0$. 
We consider this fit as our continuum limit after setting 
$N_{\tau}=\infty$, which corresponds to setting $P_2=0$ in the 
resultant fit function.
This is shown in Fig. \ref{fig:fcont}, where we see that local and 
global continuum extrapolations for $f_Q$ agree very well.

\subsection{Comparison with previous calculations}
Now let us compare the above continuum results with the previously 
published results that use the same renormalization scheme with improved 
staggered quark actions.
Namely we compare our results with the continuum results obtained with 
the stout action
\cite{Borsanyi:2010bp} as well as with the HISQ action 
\cite{Bazavov:2013yv}.
This comparison is shown in Fig. \ref{fig:fcomp}. 
We see that our results agree with the
previously published results within errors, however, the central values 
for $F_Q$ in our analysis are slightly smaller for $T<130$ MeV due to 
different way the continuum extrapolation is performed. 
The previous estimate of the continuum limit for $T \leq 135$ MeV had 
been performed by averaging $N_{\tau}=10$ and $N_{\tau}=8$ data 
\cite{Bazavov:2013yv}, whereas our analysis includes new $N_{\tau}=12$ 
ensembles at low temperatures that made a controlled continuum extrapolation 
possible. 
For $T>180$ MeV the central value of $F_Q$ in our analysis is somewhat 
larger. This is due to the updated value of the renormalization coefficients 
$c_Q$.
The previous HISQ calculations relied on the zero temperature static quark 
antiquark energies obtained in Ref. \cite{Bazavov:2011nk}, which have 
larger statistical uncertainty and use fewer $\beta$ values. 
The current analysis of $c_Q$ is based on the analysis of the zero 
temperature static quark antiquark energies from Ref. 
\cite{Bazavov:2014pvz}, which has higher statistics and uses more 
$\beta$ values.
The main new element in our analysis is that it extends to significantly 
higher temperatures.

Finally, we compare our results with the prediction of the hadron 
resonance gas (HRG) calculation for $F_Q$ \cite{Bazavov:2013yv}, which 
includes the contribution of all static-light mesons and all the 
static-light baryons (see also Ref. \cite{Megias:2012kb}). 
Since the HRG value of $F_Q$ is only defined up to a temperature 
independent constant, this constant needs to be fixed.
We do so by matching the HRG value of $F_Q$ to the lattice results at 
lowest temperature. 
The comparison is shown in Fig. \ref{fig:fcomp}.
We see that the HRG description works only for temperature $T<140$ MeV 
which is in agreement with the previous analysis \cite{Bazavov:2013yv}.
}

\section{\label{sec:entropy} Entropy of a static quark}
{
\begin{figure*}
\includegraphics[width=8cm]{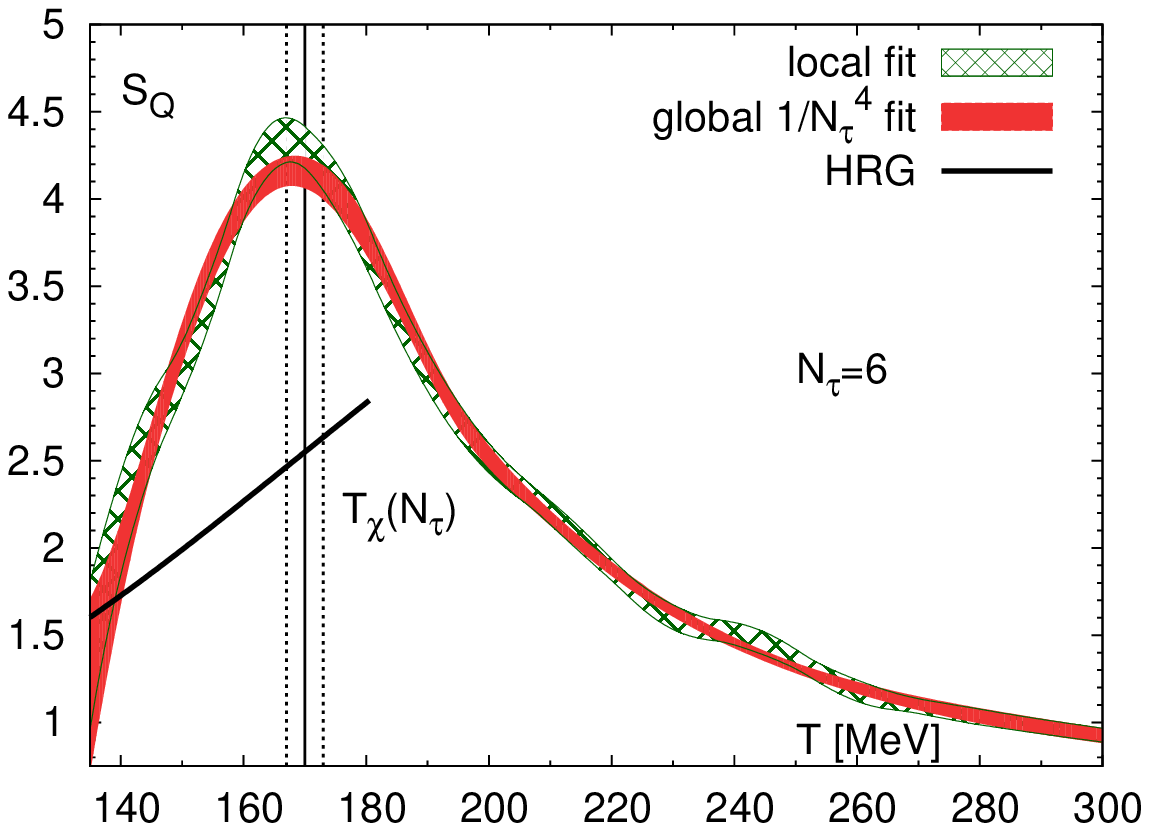}
\includegraphics[width=8cm]{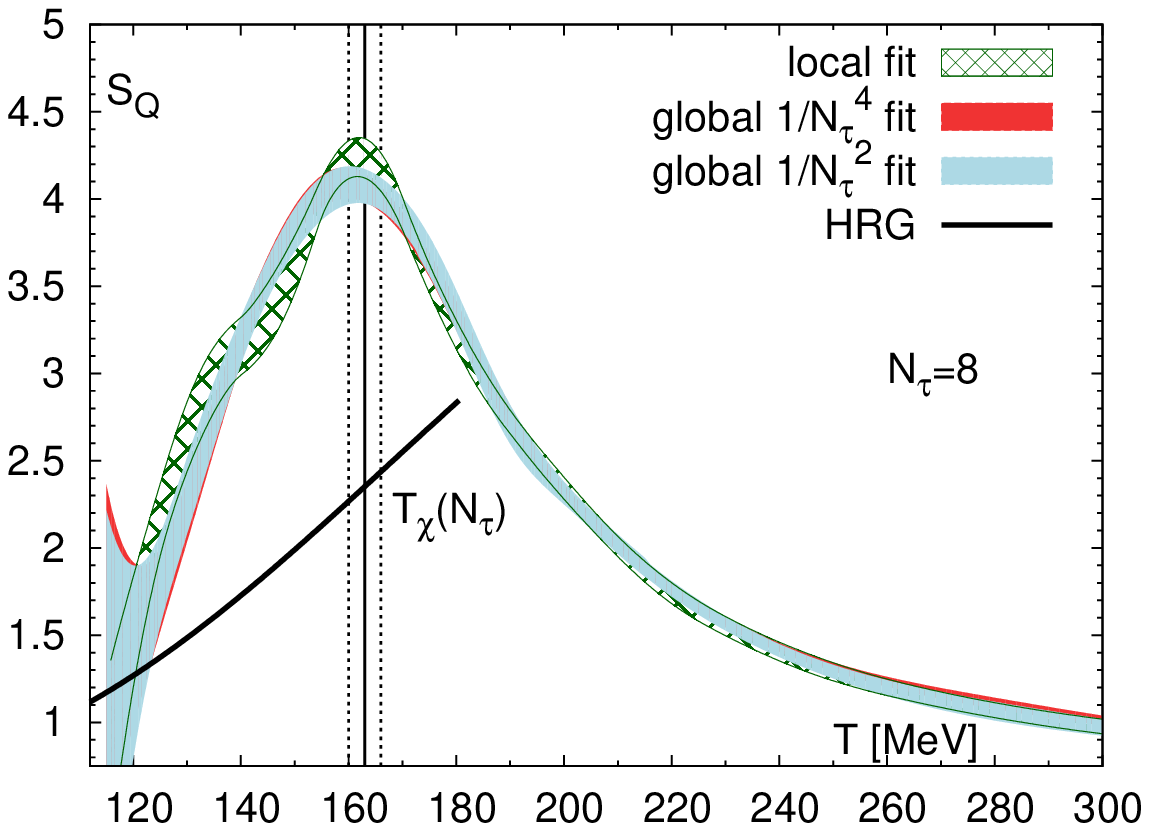}

\includegraphics[width=8cm]{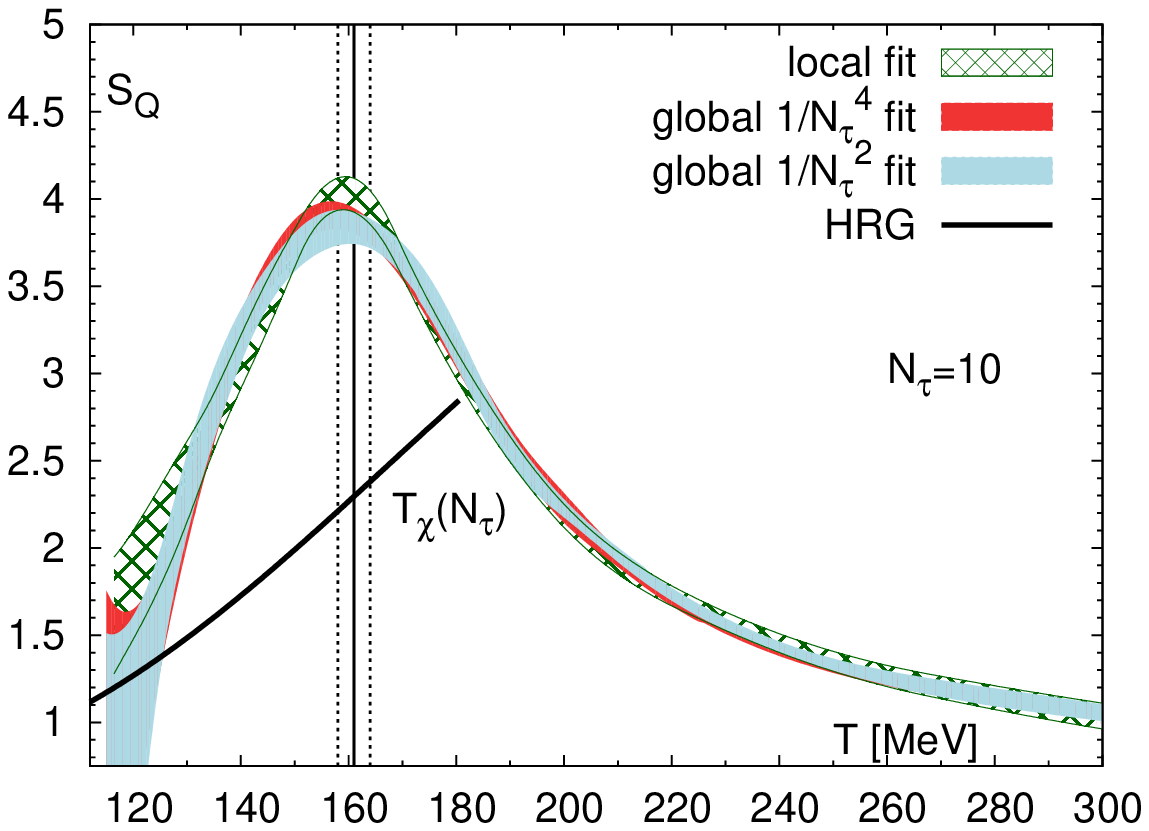}
\includegraphics[width=8cm]{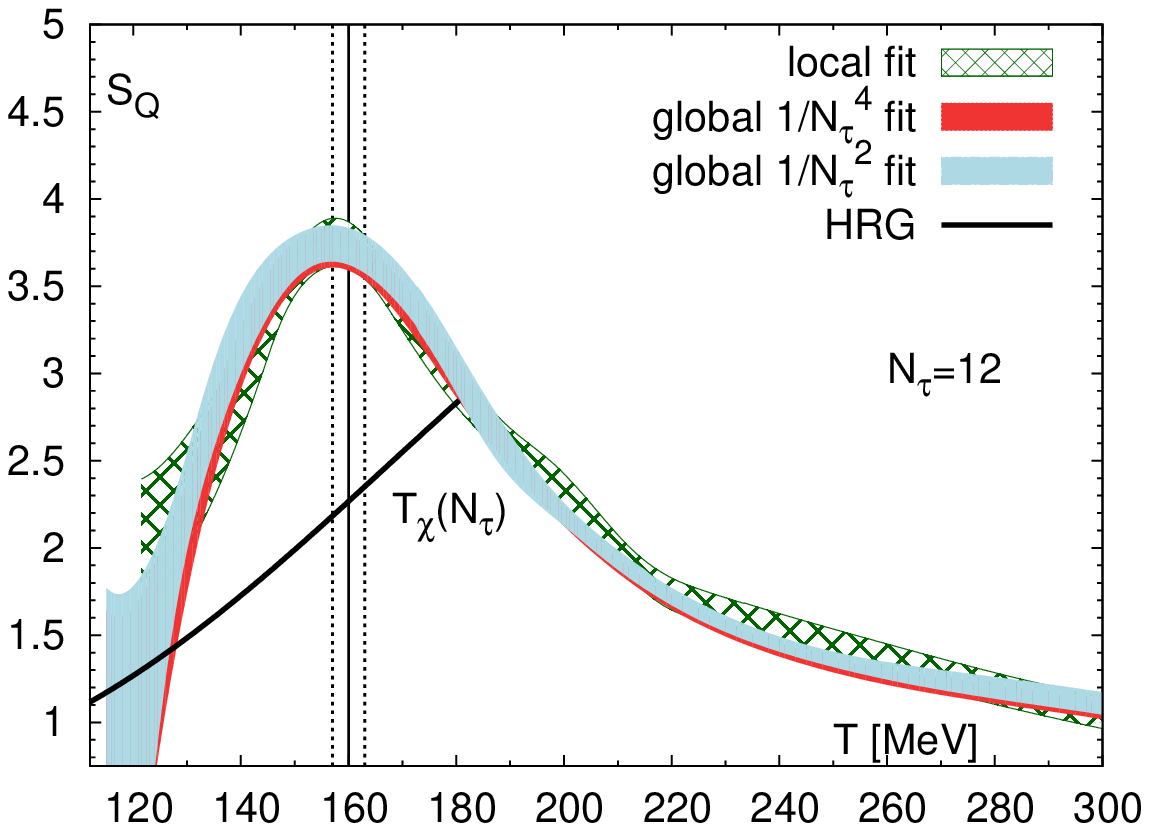}
\caption{\label{fig:SQ_allnt} 
The entropy of a static quark calculated on $N_{\tau}=6$, $8$, $10$ 
and $12$ lattices.
Shown are the results obtained from local and global fits. 
The vertical band corresponds to the chiral transition temperature 
from~\cite{Bazavov:2011nk}.
The solid black lines show the entropy in the hadron resonance gas 
model~\cite{Bazavov:2013yv}.
}
\end{figure*}
\begin{figure}
\includegraphics[width=8cm]{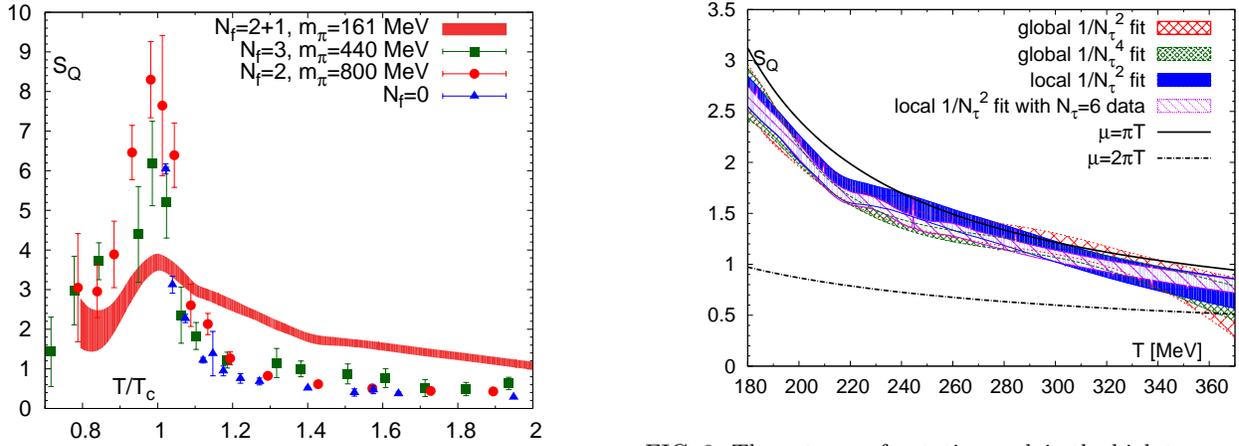}
\caption{\label{fig:sq_old_comp} 
The comparison of $S_Q$ in the continuum limit with previous calculations 
obtained on $N_{\tau}=4$ lattices\cite{Kaczmarek:2005gi, Petreczky:2004pz}. 
The temperature axis has been rescaled for each lattice calculation by a 
corresponding lattice result for $T_c$, namely 
$T_S=153\,{\rm MeV}$ for our result, $T_\chi=193\,{\rm MeV}$ and 
$T_\chi=200\,{\rm MeV}$ for the $N_f=3$ and $N_f=2$ results respectively 
and $T_L=270\,{\rm MeV}$ for the quenched case ($N_f=0$). 
}
\end{figure}
\begin{figure}
\includegraphics[width=8cm]{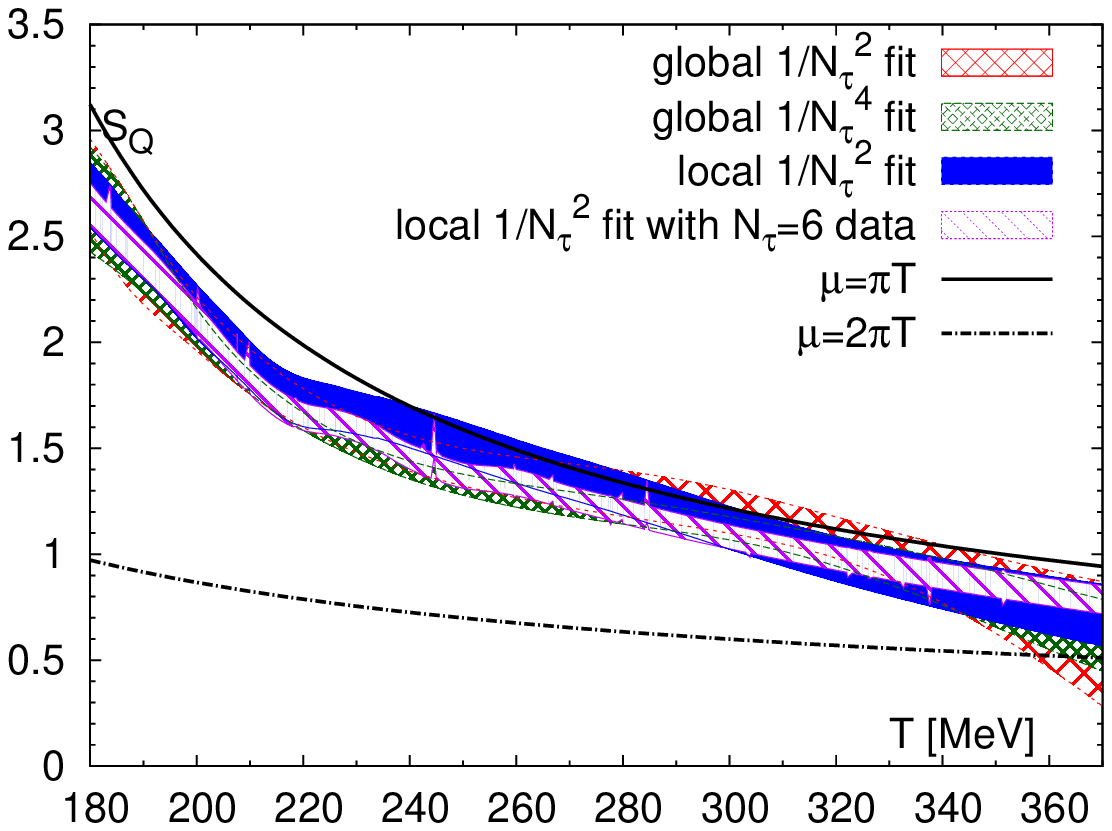}
\caption{\label{fig:SQcomp} 
The entropy of a static quark in the high temperature region. 
The lines correspond to leading order weak-coupling calculations for 
scale $\mu=2 \pi T$ and $\mu=\pi T$.
}
\end{figure}

While the free energy of a static quark encodes the screening properties 
of the hot QCD medium its temperature dependence is relatively featureless. 
The change in the screening properties of the medium can be seen more 
clearly in terms of the entropy of a static quark
\begin{equation}
S_Q(T)= -\frac{\partial F_Q(T)}{\partial T}.
\end{equation}
Note that the equality holds also if the temperature derivative is taken at 
changing volume, since the pressure exerted by a static quark is zero. 
The entropy was discussed recently in connection with the strongly coupled 
nature of quark gluon plasma \cite{Kharzeev:2014pha, Hashimoto:2014fha}.
The entropy of a static quark in SU(3) gauge theory diverges at the phase 
transition temperature and was considered in Ref. \cite{Petreczky:2005bd, 
Kaczmarek:2005gi}. 
The entropy was also calculated for 2 and 3 flavor QCD with larger than 
physical quark masses \cite{Kaczmarek:2005gi,Petreczky:2004pz}. 
It has a peak at the crossover temperature, i.e. it corresponds to the 
inflection point of $F_Q$. 
Therefore, calculating $S_Q$ for the physical quark masses is of interest, 
since $S_Q$ could be used to define a deconfinement transition temperature.

Based on the interpolation of $f_Q$ and $c_Q$ described in the previous 
section it is straightforward to estimate $S_Q$.
We write
\begin{equation}
-S_Q= f_Q^{\rm bare} + T\frac{\partial \beta}{\partial T} 
\frac{\partial f_Q^{\rm bare}}{\partial \beta}+ 
N_{\tau} ( c_Q + T\frac{\partial \beta}{\partial T} 
\frac{\partial c_Q}{\partial \beta} ).
\end{equation}
Here, the derivative $\partial \beta/\partial T$ is related to the 
nonperturbative beta function $R_\beta$ through 
$R_\beta=T( \partial \beta/\partial T)$, determined in 
Ref. \cite{Bazavov:2014pvz}. 
The entropy can  also be calculated using the global fits for 
$f_Q(T(\beta,N_\tau),N_{\tau})$ discussed in the previous section.

The numerical results for the entropy of a static quark are shown in Fig. 
\ref{fig:SQ_allnt} for $N_{\tau}=6$, $8$, $10$ and $12$ with local as well 
as global fits. 
These fits have been discussed in Secs. ~\ref{Local fits} and 
\ref{Global fits}. 
We see that with increasing temperature $S_Q$ increases reaching a maximum 
at some temperature and then decreases again.
Therefore, it makes sense to discuss the behavior of the entropy at low 
temperatures, in
the peak region and at high temperatures separately.
Since $S_Q$ for $N_{\tau}=6$ is not in the $a^2$ scaling regime in the 
peak region and below, no $ a^2 $ scaling fit is shown for $N_{\tau}=6$. 

At low temperatures we expect $S_Q$ to be described by the HRG model of 
Ref. \cite{Bazavov:2013yv}, discussed in the previous section. 
The HRG predictions from this model for $S_Q$ are shown as black lines in 
Fig. \ref{fig:SQ_allnt}. 
For low temperatures $T<130$ MeV our lattice results for $S_Q$ overlap with 
the HRG curve. 
As the temperature increases we see very clear deviations from the HRG 
result, namely the entropy $S_Q$ calculated on the lattice is significantly 
larger than the HRG prediction.

As mentioned above the entropy shows a peak at some temperature.
The position of the maximum in $S_Q$ turns out to be up to $3$ MeV below 
the chiral crossover temperature at finite cutoff, $T_\chi(N_{\tau})$ 
\cite{Bazavov:2011nk}, which is shown as a vertical line in the figure 
for each $N_{\tau}$ separately. 
The bands indicate the uncertainty in $T_\chi(N_{\tau})$.
The values of  $T_\chi(N_{\tau})$ are obtained from the $O(2)$ scaling fits 
of the chiral susceptibilities~\cite{Bazavov:2011nk}.
If the maximum in the entropy of a static quark is used to define a 
deconfinement crossover temperature one could say that deconfinement and 
chiral crossover happen at about the same temperature.

We extrapolate to the continuum with different local and global fits, 
either including a $P_4/N_{\tau}^4$ term (\mbox{cf.} Sec. \ref{Global fits}) 
and $N_{\tau}=6$ data or excluding both. 
The position of the peak scatters in the range 
$150.5$ MeV $\leq T \leq$ $157$ MeV, depending on the details of the fits, 
which are discussed in Appendix \ref{appendix:fits}. 
We consider the local fit excluding $P_4$ and $N_{\tau}=6$ as our final 
result and find the maximum of $S_Q$ at 
$T_S=153^{+6.5}_{-5}\ {\rm MeV}$. We estimate a systematic uncertainty of 
$T_S$ as $^{+4}_{-2.5}\ {\rm MeV}$ from the spread of the fits, which is 
smaller than the statistical errors that we quote.

The deconfinement transition temperature was defined as the inflection point 
of the renormalized Polyakov loop in Refs. \cite{Aoki:2006we,Aoki:2009sc} 
and values of $T_L=171(3)(4)$ MeV\footnote{
We adjusted for the change in the value of the kaon decay constant that was 
used to set the scale in Ref. \cite{Aoki:2006we} to the most recent value.} 
and $T_L=170(4)(3)$ MeV have been found, respectively.
These values are significantly larger than the chiral transition temperature. 
The most likely reason for this is that the inflection point of the 
renormalized Polyakov loop depends on the renormalization condition and 
could be different from the inflection point of $F_Q$. 
The inflection point of the renormalized Polyakov loop can be obtained 
from the equation

\begin{align}
  0 = & \frac{1}{L^{\rm ren}} 
  \frac{\partial^2 L^{\rm ren}}{\partial T^2} =
  \left(\frac{\partial f_Q}{\partial T}\right)^2 -
  \left(\frac{\partial^2 f_Q}{\partial T^2}\right) 
  \nonumber \\
  = & 
  \frac{1}{T}\left( \frac{(f_Q+S_Q)^2 -2(f_Q+S_Q)}{T} + 
  \left(\frac{\partial S_Q}{\partial T}\right) \right),
  \label{eq:scheme dependence}
\end{align}
whereas the inflection point of the free energy $F_Q$ is obtained from 
$0=\partial S_Q/\partial T$. 
In other words, the two inflection points of the Polyakov loop and the 
free energy would agree if and only if $f_Q+S_Q=0$ or $2$.
This would be the case if weak-coupling relation, $S_Q \simeq -f_Q$, was 
correct close to the crossover point.
Instead, in support of the findings in Refs. \cite{Aoki:2006we, Aoki:2009sc} 
we find the inflection point of the renormalized Polyakov loop significantly 
above the chiral transition temperature, between $180$ and $200$ MeV for 
each $N_{\tau}=12$, $10$, $8$ and $6$. 
Systematic uncertainties for $N_{\tau}=12$ are underestimated by the error 
in this range (\mbox{cf.} Appendix \ref{appendix:fits}). 
\mbox{Equation} (\ref{eq:scheme dependence}) shows that the inflection point of $L^{\rm ren}=\exp{(-f_Q)}$ depends on the term $c$ of $c_Q$ (\mbox{cf. Sec.}~\ref{sec:renL}) through $f_Q$ and $f_Q^2$. 
For $F_Q$ the change in the renormalization condition does not affect its inflection point in the continuum limit, which, in fact, does not depend on $c$.

We also compare our continuum results for $S_Q$ with previous calculations 
obtained at much larger quark masses and $N_{\tau}=4$ lattices 
\cite{Kaczmarek:2005gi, Petreczky:2004pz}.
This comparison is shown in Fig. \ref{fig:sq_old_comp}.
The temperature axis in the figure has been rescaled by the corresponding 
transition temperatures. 
We see that the peak in the entropy is much reduced compared to the previous 
calculations. 
The height of the peak is about a factor of two smaller compared to the 
previous calculations.
Both larger quark masses and fewer quark flavors correspond to physical 
settings in between QCD with 2+1 flavors at physical quark masses and pure 
gauge theory.
In pure gauge theory $S_Q$ would diverge as the temperature approaches the 
deconfinement phase transition from above. 
We further remark that Fig.~\ref{fig:SQ_allnt} clearly shows that the height 
of the peak decreases for increasing $N_{\tau}$. 
Therefore, one would generally expect to see a higher peak in $S_Q$ at 
finite cutoff than in the continuum limit. 
Hence, the much reduced height of the peak is no surprise. 

Finally, let us discuss the behavior of $S_Q$ in the high temperature region. 
For $T>220$ MeV we have sufficiently accurate data for all lattice spacings. 
We have performed several continuum extrapolations based on global and local 
fits. 
These are shown in Fig. \ref{fig:SQcomp}. 
We can see from the figure that different continuum extrapolations have 
overlapping error bands. 
In particular $N_\tau=6$ data is consistent with $1/N_{\tau}^2$ scaling 
behavior. 
The uncertainty grows significantly, however, as we approach $T=400$ MeV 
due to the fact that renormalization constants are available only up to that 
temperature for $N_{\tau}=12$ data. 
In the next section we will discuss how to extend the results to higher 
temperatures. 
In Fig. \ref{fig:SQcomp} we also show the results for weak-coupling 
calculations at leading order with one-loop running coupling for two 
different renormalization scales. 
As one can see from the figure the LO result for $S_Q$ is not very 
different from the lattice calculations, however, the scale dependence 
is quite large. 
Furthermore, higher order corrections are also important. 
Therefore, for a meaningful comparison of the lattice and the 
weak-coupling results it is necessary to extend the calculations to 
higher temperatures and to higher orders in the perturbative expansion. 
This will be discussed in Sec. \ref{sec:weak}.
}

\section{\label{sec:high} Polyakov loop at high temperatures}
{
The highest temperature at which we can study the Polyakov loop
or equivalently $F_Q$ so far was limited by the knowledge of $c_Q$
determined by the zero temperature static $Q\bar Q$ energy.
Below we will discuss a method to work around this limitation which we
call the direct renormalization scheme.

The idea of the direct renormalization scheme is to determine $c_Q$ by 
comparing the free energy $f_Q$ calculated for the same temperature but 
different $N_\tau$ \cite{Gupta:2007ax}. 
\mbox{Equation}~(\ref{eq: fqren}) can be applied to obtain $c_Q(\beta)$ once 
$f_Q(T(\beta,N_{\tau}),N_{\tau})$ and $f_Q^{\rm{bare}}(\beta,N_\tau)$ 
are known. 
If there were no cutoff effects in $f_Q(T(\beta,N_\tau),N_{\tau})$ after 
renormalization, $c_Q(\beta)$ at some value of $\beta$ would read

\begin{align}
  c_Q(\beta) =&\ \frac{1}{N_{\tau}} 
  \big[ N_{\tau}^{\rm{ref}} c_Q(\beta^{\rm{ref}}) + \nonumber\\
  &\ f_Q^{\rm{bare}}(\beta^{\rm{ref}},N_{\tau}^{\rm{ref}})-
  f_Q^{\rm{bare}}(\beta,N_{\tau}) \big],
\end{align}
where $N_\tau^{\rm ref}$ and $\beta^{\rm ref}$ correspond to a reference 
point, where $c_Q$ is known.

\begin{figure}[b]
\includegraphics[width=7cm]{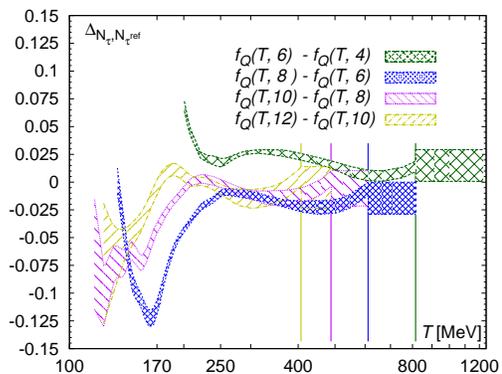}
  \caption{\label{fig:Delta} 
  Extrapolation of residual cutoff effects in 
  $ f_Q^{\rm{ren}}(T(\beta,N_{\tau}),N_\tau^{\rm{ref}})$. 
  The vertical lines indicate the start of the extrapolated 
  $ \Delta_{N_\tau,N_\tau^{\rm{ref}}}(T) $ for each pair 
  $( N_\tau,N_\tau^{\rm{ref}} )$.
  }
\end{figure}

\begin{figure}
\includegraphics[width=7cm]{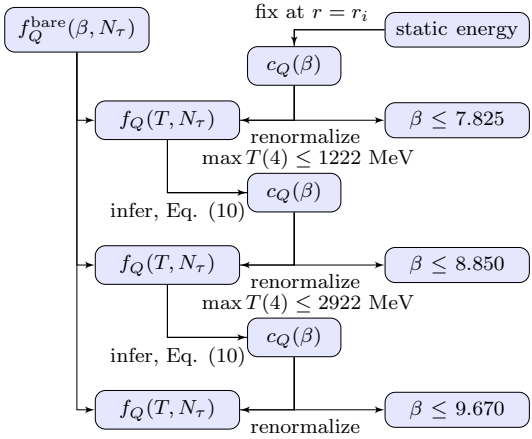}
\caption{\label{fig:flowchart} 
The flow chart sketches the different steps of the direct renormalization procedure. 
For each step the temperature $T(\beta,N_\tau)$ is limited by the corresponding $\beta \leq \beta_{\max}$.
}
\end{figure}

Next, we study the cutoff dependence of $f_Q$. 
It is convenient to do so by considering the following difference

\begin{align}
  \Delta_{N_\tau,N_\tau^{\rm{ref}}}(T) 
  =& f_Q(T(\beta,N_{\tau}),N_{\tau}) \nonumber \\
  -& f_Q(T(\beta^{\rm ref},N_{\tau}^{\rm ref}),N_\tau^{\rm{ref}}).
  \label{eq: residual cutoff effects}
\end{align}
In Fig. \ref{fig:Delta} we show $\Delta_{N_\tau,N_\tau^{\rm{ref}}}(T)$ 
as function of the temperature for different combinations of $N_\tau$ 
and $N_\tau^{\rm ref}$.
At low temperatures, $T<250$ MeV, this quantity shows a strong 
temperature dependence. 
However, for $T>250$ MeV the temperature dependence of 
$\Delta_{N_\tau,N_\tau^{\rm{ref}}}(T)$ is rather mild, and one may 
approximate it by a constant.
Therefore, we assume that above the temperatures where no lattice data 
for $c_Q$ are available $\Delta_{N_\tau,N_\tau^{\rm{ref}}}(T)$ is constant. 
If predictions for $c_Q$ from all possible pairs $(N_\tau,N_{\tau}^{\rm{ref}})$ 
are consistent within uncertainties, one may conclude in retrospect that the 
assumption was justified. 
We estimate its central value from the average of the minimum and the maximum 
of the one sigma band of 
$\Delta_{N_\tau,N_\tau^{\rm{ref}}}(T)$ for $ T>250 $ MeV and its uncertainty 
by the respective difference.
This estimate is shown in Fig. \ref{fig:Delta}.
Using $\Delta_{N_\tau,N_\tau^{\rm{ref}}}^{\rm av}(T)$ determined this way 
together with the corresponding error we can provide an estimate for $c_Q$ 
that should be free of cutoff effects:

\begin{align}
  c_Q(\beta) =&\ \frac{1}{N_{\tau}} 
  \big[ N_{\tau}^{\rm{ref}} c_Q^{Q \bar Q}(\beta^{\rm{ref}}) + 
  \Delta_{N_\tau,N_\tau^{\rm{ref}}}^{\rm av} + \nonumber\\
  &\ f_Q^{\rm{bare}}(\beta^{\rm{ref}},N_{\tau}^{\rm{ref}})-
  f_Q^{\rm{bare}}(\beta,N_{\tau}) \big] .
  \label{eq: direct renormalization}
\end{align}

\begin{figure}[b]
\includegraphics[width=8cm]{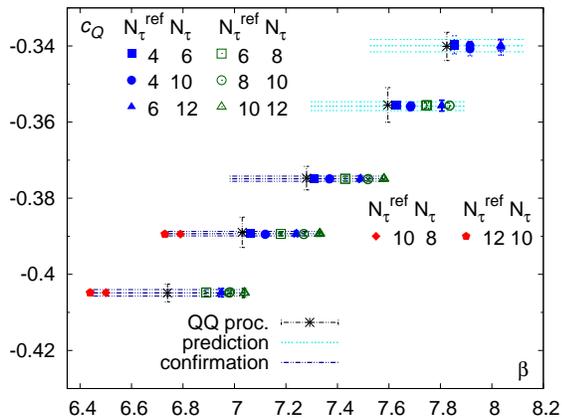}
  \caption{\label{fig: comparehizren} 
  Comparison between renormalization constant $ c_Q(\beta) $ from direct 
  renormalization and $ Q \bar Q $ procedures. 
  Symbols and data are explained in the text. 
  }
\end{figure}
We use all possible pairs $( N_\tau,N_\tau^{\rm{ref}} )$ and compute 
$ c_Q^{\rm{direct}}(\beta,N_\tau,N_\tau^{\rm{ref}}) $ via 
\mbox{Eq.}~(\ref{eq: direct renormalization}) from $ c_Q^{Q \bar Q}(\beta) $ 
for all possible temperatures. 
We can only calculate $c_Q$ with direct renormalization procedure up to 
$\beta=8.57$, if we use $N_{\tau}=8$ results for the bare Polyakov loops 
($T(\beta=8.57,N_{\tau}=8)=1155$ MeV)
or to $\beta=8.85$ if we use $N_{\tau}=12$ results for the bare Polyakov loop 
($T(\beta=8.85,N_{\tau}=12)=974$ MeV). 
To extend the beta range even further, we use the two step procedure for the 
direct renormalization. 
First, we compute $ c_Q^{\rm direct} $ up to $ \beta=8.85 $ from 
$c_Q^{Q \bar Q}$ in the first iteration. 
Next, we add the new values of the renormalization constant to the bare 
free energies up to $ T(\beta=8.85,4)=2922 $ MeV. 
Finally, we compute $ c_Q^{\rm direct} $ up to $ \beta=9.67 $ from 
$ c_Q^{\rm direct} $ in a second iteration and add the new values of the 
renormalization constant to bare free energies up to 
$ T(\beta=9.67,4)=5814 $ MeV. 
We sketch the procedure in the flow chart in Fig.~\ref{fig:flowchart}.
In order to test robustness and predictive power of direct renormalization, 
we omit $ c_Q^{Q \bar Q}(\beta) $ for $ \beta>7.373 $
and calculate $c_Q^{\rm direct}$ using the above procedure.
After excluding $c_Q^{Q \bar Q}(7.596)$  and $ c_Q^{Q \bar Q}(7.825) $ from 
the input, we compare 
the predictions for $ c_Q^{\rm{direct}}(7.596,N_\tau,N_\tau^{\rm{ref}}) $ 
and $ c_Q^{\rm{direct}}(7.825,N_\tau,N_\tau^{\rm{ref}}) $ with known values 
of $ c_Q^{Q \bar Q}(\beta) $. 
We show this comparison for a few selected $ \beta $ values and pairs 
($ N_\tau,N_\tau^{\rm{ref}} $) in \mbox{Fig.}~\ref{fig: comparehizren}. 
Black bursts represent $ c_Q^{Q \bar Q}(\beta) $ data from zero temperature 
lattices.
Results $ c_Q^{\rm{direct}}(\beta,N_\tau,N_\tau^{\rm{ref}}) $ inferred from 
coarser \mbox{resp.} finer lattices ($ N_\tau>N_\tau^{\rm{ref}} $ 
\mbox{resp.} $ N_\tau<N_\tau^{\rm{ref}} $) are displaced to the right 
\mbox{resp.} left of $ c_Q^{Q \bar Q}(\beta) $. 
Shape and color of the symbols encode $ N_\tau^{\rm{ref}} $ and $ N_\tau $.  
As one can see from the figure the direct renormalization method correctly 
reproduces the values of the renormalization constant obtained in the 
$Q\bar Q$ procedure.

Since no trends in $ c_Q^{\rm{direct}}(\beta,N_\tau,N_\tau^{\rm{ref}}) $ 
depending on either $ N_\tau $ or $ N_\tau^{\rm{ref}} $ are observed, we 
conclude that no residual cutoff effects are present. 
We average over all possible pairs $( N_\tau,N_\tau^{\rm{ref}} )$ that 
reproduce one of the $ \beta $ values of an underlying Polyakov loop within 
$ \pm 0.01 $, take the error's mean as statistical error and the standard 
deviation as systematical error estimate (at most 25\% of the statistical 
error). 
We add these errors in quadrature and show the $ 1\sigma $ bands of 
$ c_Q^{\rm{direct}}(\beta) $ in the figure. 
We show with dark blue lines (for $ \beta \leq 7.373 $) that input values 
$ c_Q^{Q \bar Q}(\beta) $ are reproduced.
Hence, consistency between both renormalization schemes is evident. 
We show with cyan lines (for $ \beta > 7.373 $) 
that predictions of the direct renormalization procedure 
are consistent with $ c_Q^{Q \bar Q}(\beta) $ outside of the input 
$ \beta $ range. 
Therefore, we confirm that our approach for the direct renormalization 
procedure has predictive power outside of the 
input $ \beta $ range 
and that our extrapolation assuming constant cutoff effects in 
\mbox{Fig.}~\ref{fig:Delta} is justified. 
\begin{figure}[t]
\includegraphics[width=7cm]{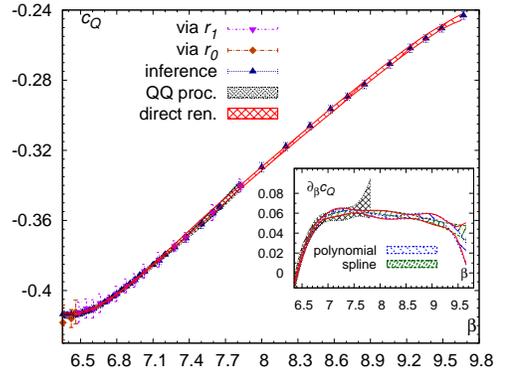}
  \caption{\label{fig: directzren} 
  Renormalization constant $ c_Q(\beta) $ from direct renormalization 
  and $ Q \bar Q $ procedures. 
  Interpolations are shown as $ 1\sigma $ bands and data points are 
  explained in the text. 
  The inset shows the derivative $\frac{\partial c_Q}{\partial \beta}$.
  }
\end{figure}

\begin{figure*}
\includegraphics[width=8cm]{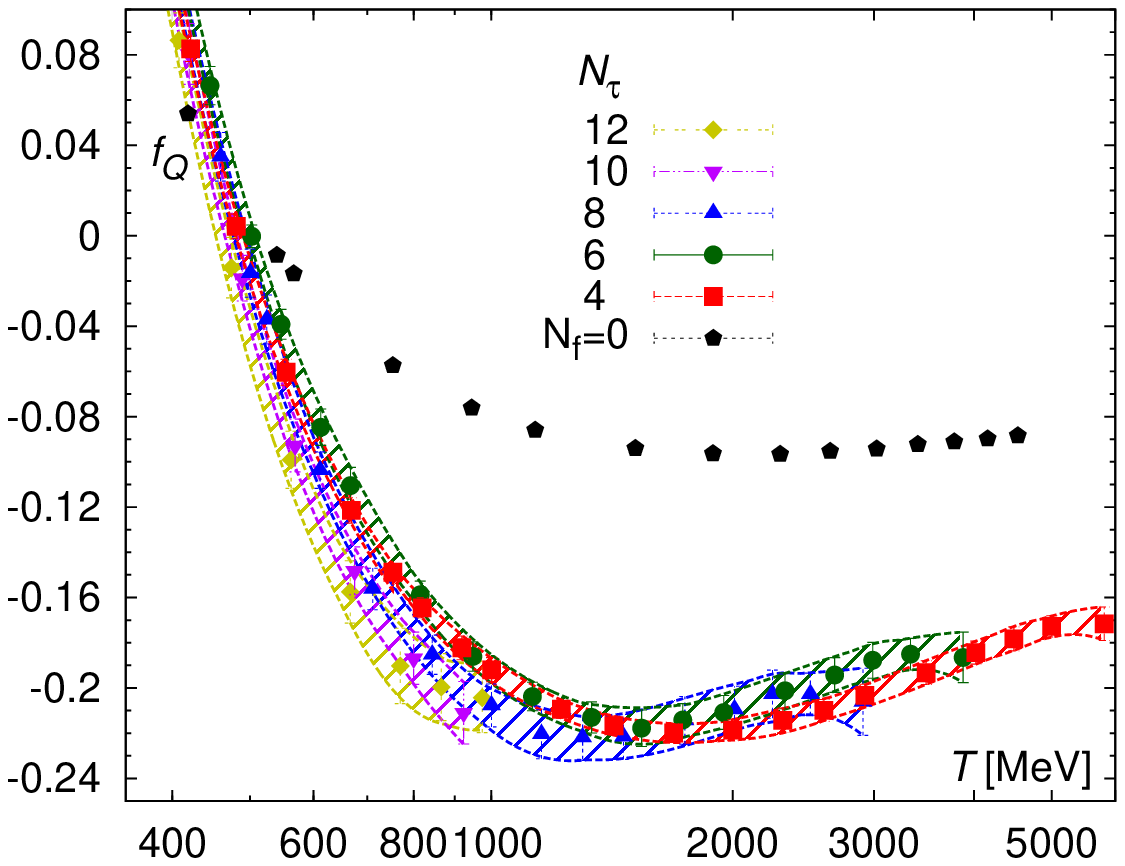}
\includegraphics[width=8cm]{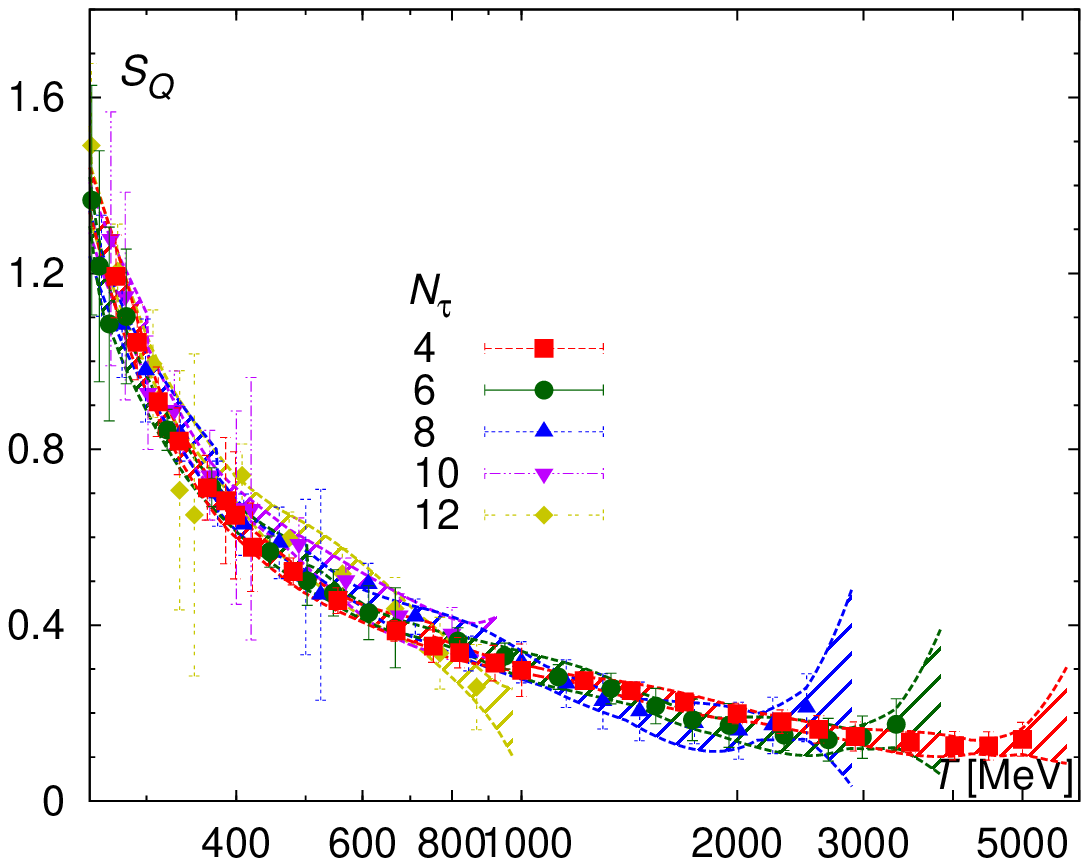}
\caption{\label{fig:high_FQ_SQ} 
The free energy (left) and the entropy (right) of a static quark in the 
high temperature region. 
The bands show the results of interpolation with the corresponding 
uncertainty. 
For comparison, the free energy for SU(3) Yang-Mills theory on $ N_\tau=4 $ 
lattices is included~\cite{Gupta:2007ax}.
The symbols on the right correspond to $S_Q$ calculated from finite differences.
}
\end{figure*}
Having determined the renormalization constant in the extended range of $\beta$ 
(\mbox{cf.} \mbox{Fig.}~\ref{fig: directzren}) it is straightforward to 
calculate the free energy $f_Q$ at considerably higher temperatures. 
Namely our calculations with $N_{\tau}=12$ now extend to $T=900$ MeV, while 
for $N_{\tau}=6$ and $N_{\tau}=8$ we can reach to temperatures of about 
$3800$ MeV and $2900$ MeV, respectively.
The results of our calculations at high temperatures ($T>400$ MeV) are shown 
in Fig. \ref{fig:high_FQ_SQ} for different $N_{\tau}$. 
In the figure we also show the local interpolation of the data as bands. 
One can see that the cutoff dependence of the data is rather mild, i.e. the 
bands corresponding to different $N_{\tau}$ are largely overlapping, 
including the $N_{\tau}=4$ results. 
In other words, even for our coarsest lattice the cutoff effects are very 
small in this high temperature region. 
This will be important for the comparison with the weak-coupling calculations 
discussed in Sec. \ref{sec:weak} since this comparison can be performed 
using the $N_{\tau}=4$ results that extend up to temperatures as high as 
$5814$ MeV. 
We also note that the free energy becomes negative for $T>500$ MeV as 
expected from the weak-coupling calculations. 
The other interesting feature of $f_Q$ is that it has a minimum around 
temperatures of about $1500$ MeV corresponding to a maximum of the 
renormalized Polyakov loop. 
This feature was observed in the SU(3) gauge theory, where the renormalized
Polyakov loop has the maximum at temperatures of $12T_d$, with $T_d$ 
being the deconfinement phase transition temperature \cite{Gupta:2007ax}. 
These SU(3) Yang-Mills theory results, which have been included in 
\mbox{Fig.}~\ref{fig:high_FQ_SQ}, yield significantly smaller $|f_Q|$ than 
our results with 2+1 flavors. 
The difference is pronounced most strongly in the vicinity of the minimum 
of $f_Q$.

From the interpolations of $f_Q$ it is straightforward to calculate the 
entropy
of a static quark. This is shown in Fig. \ref{fig:high_FQ_SQ} (right panel). 
Furthermore, since for $T>400$ MeV the free energy varies smoothly with the 
temperature it is possible to calculate $S_Q$ without any interpolation. 
We could estimate $S_Q$ by approximating the temperature derivative of 
$F_Q$ by finite differences of the lattice data on $F_Q$ at two 
neighboring temperature values. 
The entropy estimated from the finite differences is also shown in 
Fig. \ref{fig:high_FQ_SQ} and it agrees very well with the results obtained 
from interpolations.
For $T>900$ MeV we have $S_Q \simeq -f_Q$ as expected in the 
weak-coupling picture.
We also note that the entropy at high temperatures is also higher than 
in the SU(3) gauge theory.
}

\section{\label{sec:flow} Renormalization with gradient flow}
{
\begin{figure}
\includegraphics[width=8cm]{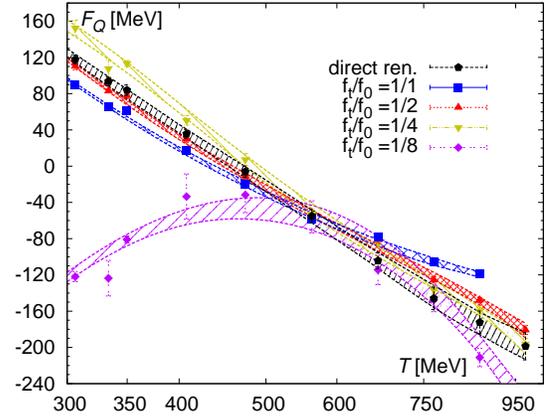}
\caption{ 
The free energy of a static quark calculated on $N_{\tau}=12$ lattices 
for different flow times.
}
\label{fig:nt12flow}
\end{figure}
\begin{figure}
\includegraphics[width=8cm]{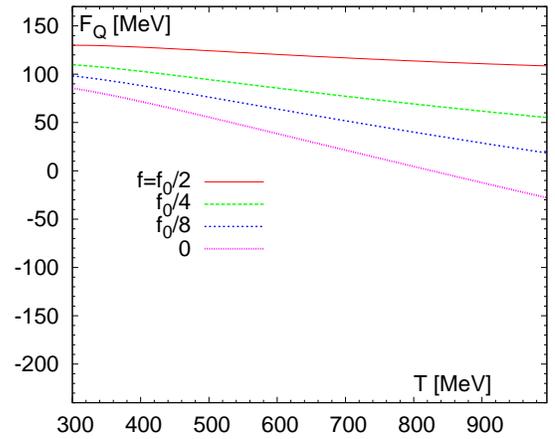}
\caption{ 
The free energy of a static quark at leading order calculated for 
different flow times. 
The values of the free energy have been shifted by $300$ MeV (see text).
}
\label{fig:LO_fdep}
\end{figure}

The gradient flow was introduced as a tool to remove short distance 
divergences in the lattice observables 
\cite{Luscher:2010iy, Luscher:2011bx}. 
It is defined by the differential equation \cite{Luscher:2010iy}
\begin{equation}
\frac{d V_{\mu}(x,t)}{d t}=-g_0^2 \partial_{x,\mu} S[V] V_{\mu}(x,t), 
\label{flow}
\end{equation}
where $S[V]$ is the lattice gauge action and $g_0^2=10/\beta$ is the 
bare lattice gauge coupling. 
The new link variable $V_{\mu}(x,t)$ has the initial value given by 
the original link variable $V_{\mu}(x,t=0)=U_{\mu}(x)$.
Here we use the same notation for $\partial_{x,\mu} S[V]$ as in 
Ref. \cite{Luscher:2010iy}.
The gradient flow has been extensively used at zero temperature for scale 
setting (see, e.g., Ref. \cite{Borsanyi:2012zs,Bazavov:2015yea}) as well 
as at nonzero temperature for the calculations of the equation of 
state \cite{Asakawa:2013laa}.
In Ref. \cite{Petreczky:2015yta} it was proposed to use the gradient flow 
to calculate the renormalized Polyakov loops. 
It was shown there that up to a temperature independent constant the free 
energy of a static quark calculated using the gradient flow agrees with 
the free energy obtained in the conventional ($Q\bar Q$) scheme in the 
continuum limit up to temperatures $T=400$ MeV, provided that the flow 
time $f=\sqrt{8 t}$ satisfies the condition:
\begin{equation}
a \ll f \ll 1/T,~~{\rm or}~~~ 1 \ll f T \ll N_{\tau}.
\label{eq:scaling_cond}
\end{equation}
The gradient flow method also enabled the calculation of the free energy 
of static charges in higher representation and confirmed the expected 
Casimir scaling in the high temperature region \cite{Petreczky:2015yta}.
Here we would like to extend these studies to higher temperatures and 
also analyze the fluctuations of the Polyakov loop.

\subsection{Renormalized Polyakov loop from gradient flow}

We followed the procedure outlined in Ref. \cite{Petreczky:2015yta} and 
calculated the Polyakov loop at nonzero flow time by replacing the 
link variables $U_{\mu}(x)$ in Eq. (\ref{defP}) by $V_{\mu}(x,t)$. 
We use the tree level Symanzik gauge action in Eq. (\ref{flow}).
We calculated the Polyakov loop for the same flow times as in 
Ref. \cite{Petreczky:2015yta}, namely, 
$f=\sqrt{8 t}=f_0,~3/4 f_0, 1/2 f_0, 1/4 f_0$ and $1/8 f_0$, 
$f_0=0.2129$ fm.
See Ref. \cite{Petreczky:2015yta} for further details. 
In Fig. \ref{fig:nt12flow} we show our numerical results for 
$N_{\tau}=12$ shifted by a constant such that the results obtained 
at different flow times agree with the continuum result for $F_Q$ 
obtained in the previous section at $T=600$ MeV. 
The bands shown in the figure correspond to the interpolation of 
the lattice data.
One can see from the figure that the temperature dependence of $F_Q$ 
obtained with $f=f_0,~3/4 f_0, 1/2 f_0$ is very similar to the 
temperature dependence of the free energy obtained using the direct 
renormalization procedure for $T<500$ MeV. 
With a suitable constant shift all these results can be made to agree 
with each other in this temperature region. 
For higher temperatures, however, the temperature dependence of $F_Q$ 
obtained with these values of the flow time is not captured correctly. 
Choosing a smaller flow time, namely $f=f_0/4$, the temperature dependence 
of $F_Q$ obtained using direct renormalization method is reproduced. 
However, decreasing the flow time even further to $f_0/8$ leads to a 
completely different temperature dependence. 
Thus, for $T>500$ MeV the results are very sensitive to the choice of 
the flow time, i.e. the scaling window is very narrow. 
We also performed the calculations for $N_{\tau}=6,~8$ and $10$. 
The corresponding results are similar to the ones shown in 
Fig. \ref{fig:nt12flow} but the flow time dependence is even stronger. 
This stronger flow time dependence is expected 
(cf. Eq. (\ref{eq:scaling_cond})).

\begin{figure*}
\includegraphics[width=8cm]{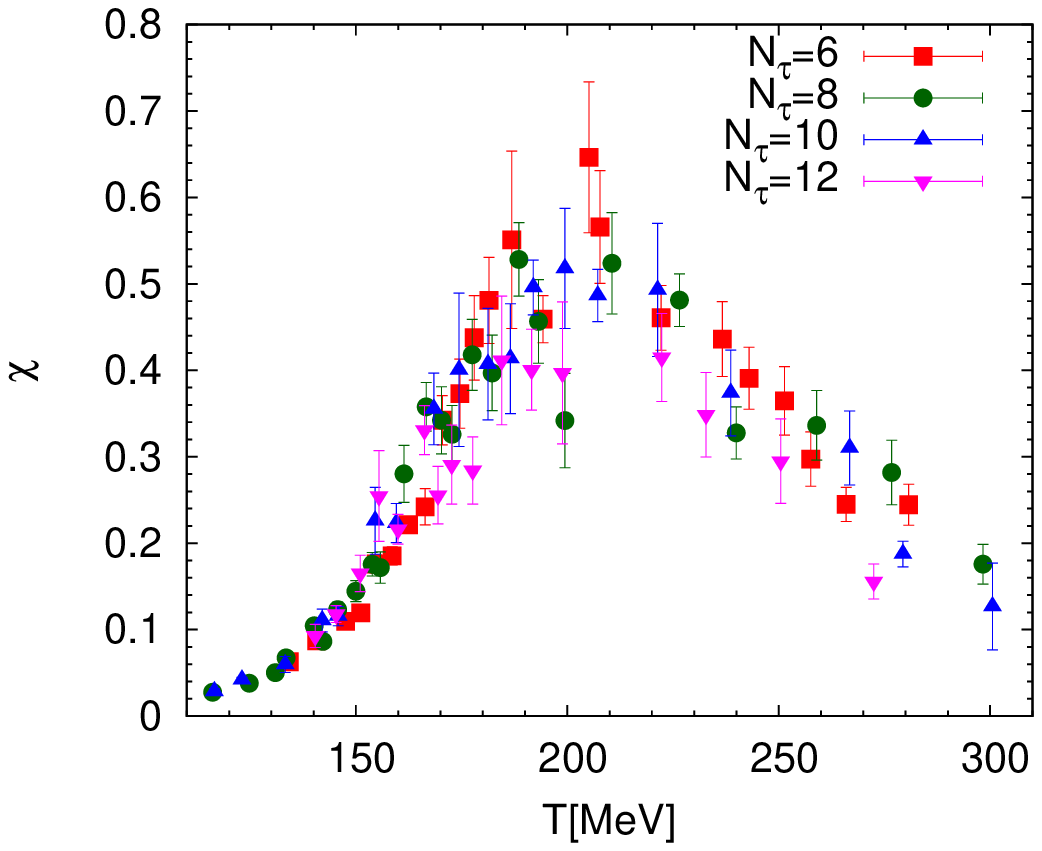}
\includegraphics[width=8cm]{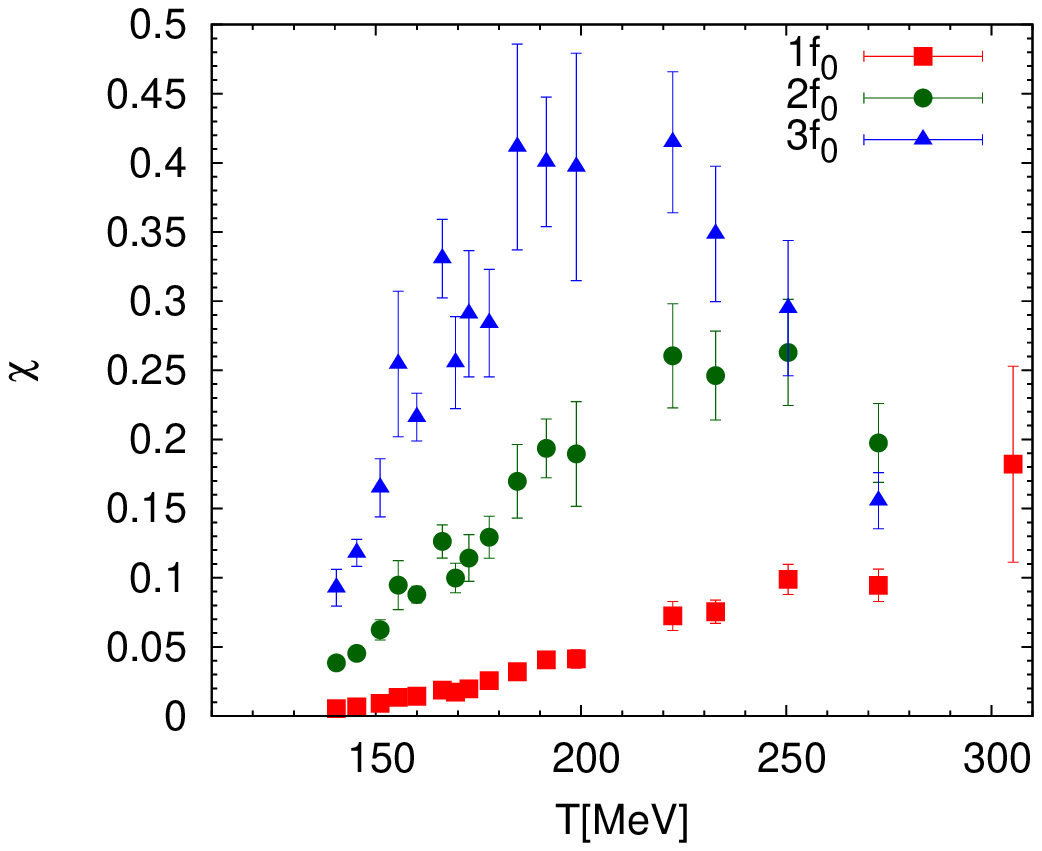}
\caption{\label{fig:chi} 
The Polyakov loop susceptibility obtained using gradient flow 
for $f=3f_0$ and
different $N_{\tau}$ (left) and for $N_{\tau}=12$ and 
$f=f_0, 2f_0$ and $3 f_0$.}
\end{figure*}
To understand the flow time dependence of the free energy of a 
static quark shown
in Fig. \ref{fig:nt12flow} it is useful to analyze the leading order 
result for
the Polyakov loop obtained at nonzero flow time \cite{Datta:2015bzm}.
In terms of the free energy the leading order result reads
\begin{equation}
F_Q^f(T)=C_F \alpha_s \frac{\sqrt{\pi}}{f}-
C_F \alpha_s \frac{m_D}{2} \tilde \Phi(m_D f/2),
\label{LO_f}
\end{equation}
where 
$\displaystyle \tilde \Phi(z)=e^{z^2} 
\frac{2}{\sqrt{\pi}} \int_z^{\infty} d x e^{-x^2}$. 
Here and in what follows we use the label $f$ on the free energy to 
denote the free energy obtained with gradient flow.
For sufficiently small flow time this result approaches the well 
known leading order result for $F_Q$ (up to a temperature independent 
constant $\sim 1/f$), since $\tilde \Phi(z=0)=1$. 
Now the question arises which value of the flow time can be considered 
as sufficiently small. 
Therefore, in Fig. \ref{fig:LO_fdep} we show the leading order result 
given by Eq. (\ref{LO_f}) omitting the constant term $\sim 1/f$. 
Furthermore, we shifted $F_Q^f$ by $300$ MeV to facilitate the comparison 
with the lattice results. 
We see a similar trend in the flow time dependence of the leading order 
result for
$F_Q^f(T)$: 
As the flow time increases the temperature dependence becomes milder. 
For $T<400$ MeV $f=f_0/4$ can be considered as sufficiently small. 
However, at higher temperature we must have $f< f_0/8$. 
On the other hand, as we have seen above, the value of $f=f_0/8$ is too 
small for $N_{\tau}=12$ lattices to remove the lattice artifacts. 
This suggests that one has to use lattices with temporal extent 
$N_{\tau}>12$ to obtain the correct temperature dependence of the Polyakov 
loop for $T>400$ MeV.

One could also try to follow a different philosophy and fix the flow time 
such that $f \cdot T=const.$ as it was done in Ref. \cite{Datta:2015bzm}. 
In this case the term proportional to $1/f$ would contribute to the 
temperature dependence of $F_Q^f$ and thus to the entropy 
$S_Q^f=-\partial F_Q^f/\partial T$. 
The additional contribution to the entropy just amounts to a constant shift 
compared to the entropy of a static charge defined in the conventional way, 
i.e. the temperature dependence of the entropy would be the same as before. 
By matching the entropy obtained from the gradient flow to the entropy of a 
static quark obtained in the conventional scheme one could in principle 
obtain results for the entropy at higher temperatures. 
We tried to implement this scheme, however, it turns out that the resulting 
errors are too large to obtain reliable results for the entropy of a static 
charge at high temperatures. 

\begin{figure*}
\includegraphics[width=5.8cm]{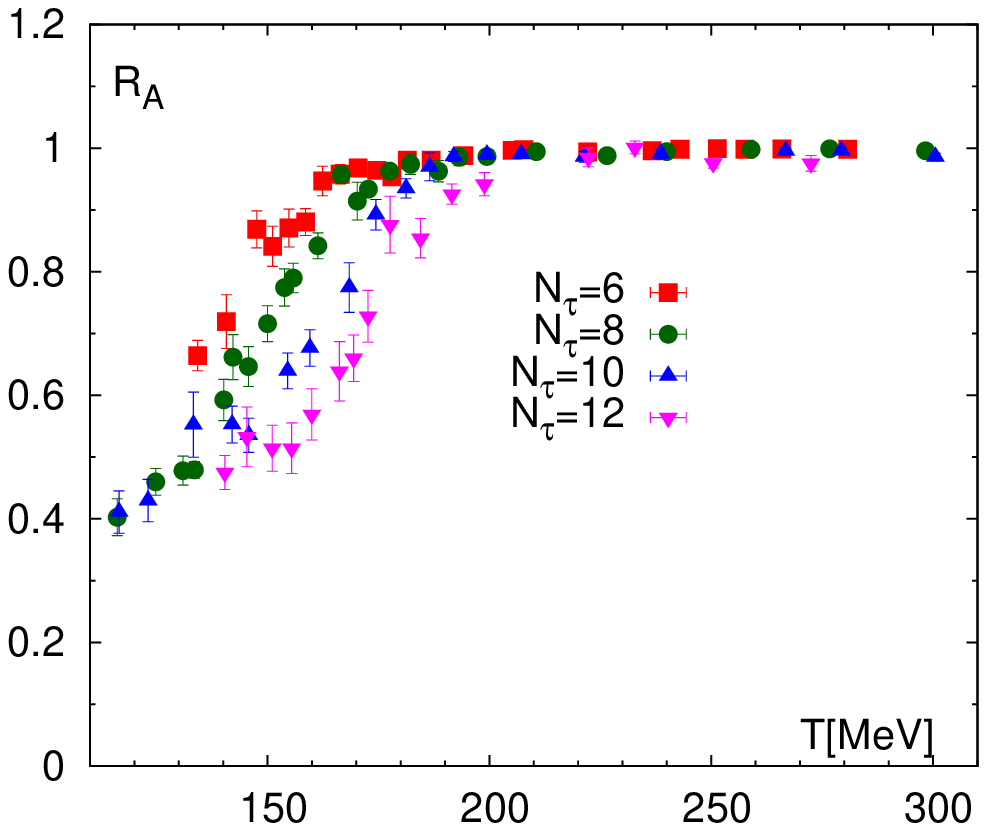}
\includegraphics[width=5.8cm]{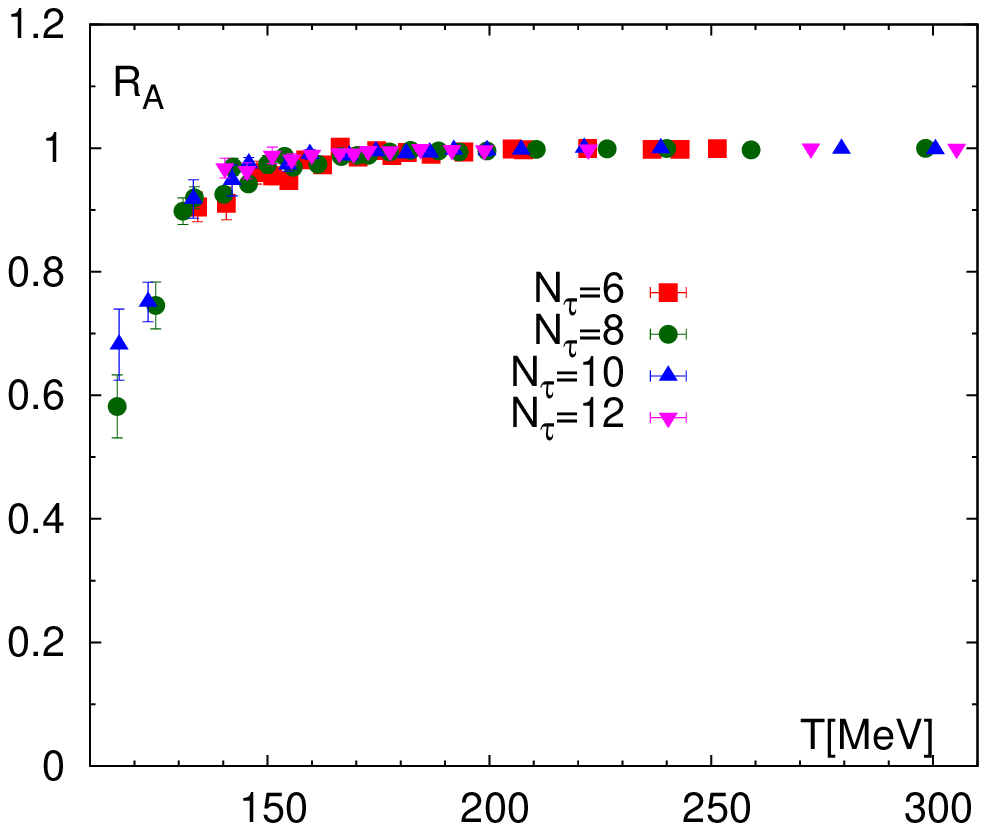}
\includegraphics[width=5.8cm]{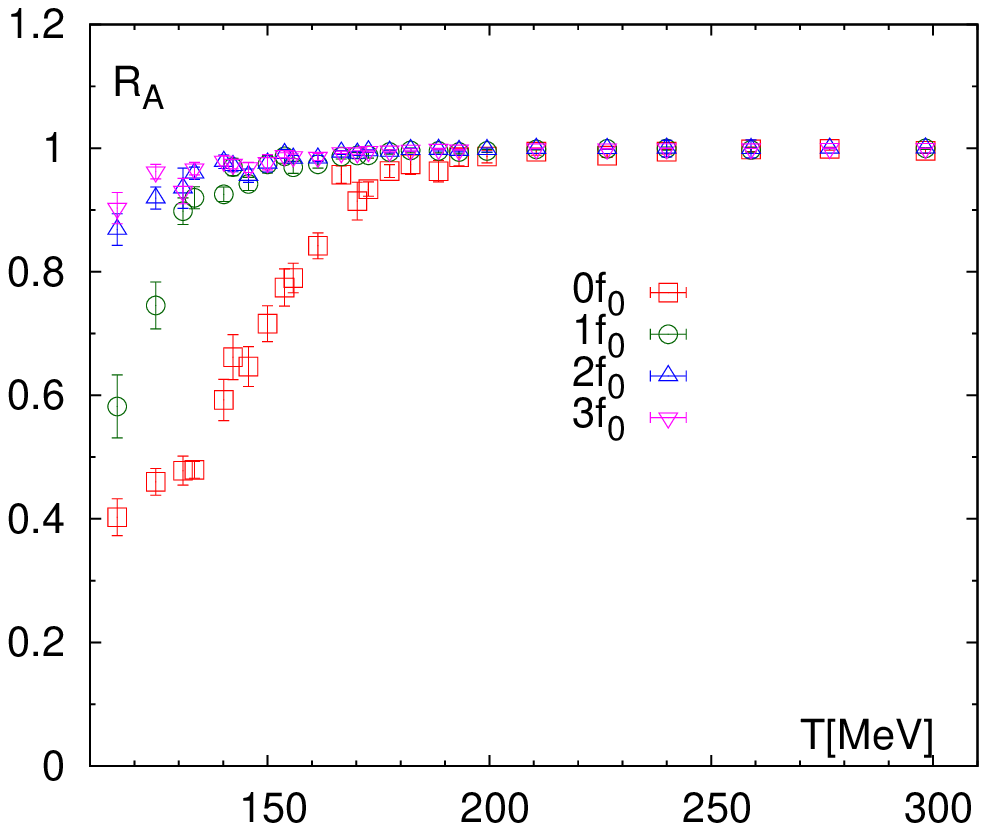}
\caption{\label{fig:ratA} 
The ratio of the susceptibilities $R_A$ shown as function of the temperature
for zero flow time (left), flow time $f=f_0$ (middle) and for different flow 
times but for $N_{\tau}=8$ (right).}
\end{figure*}
\begin{figure*}
\includegraphics[width=5.8cm]{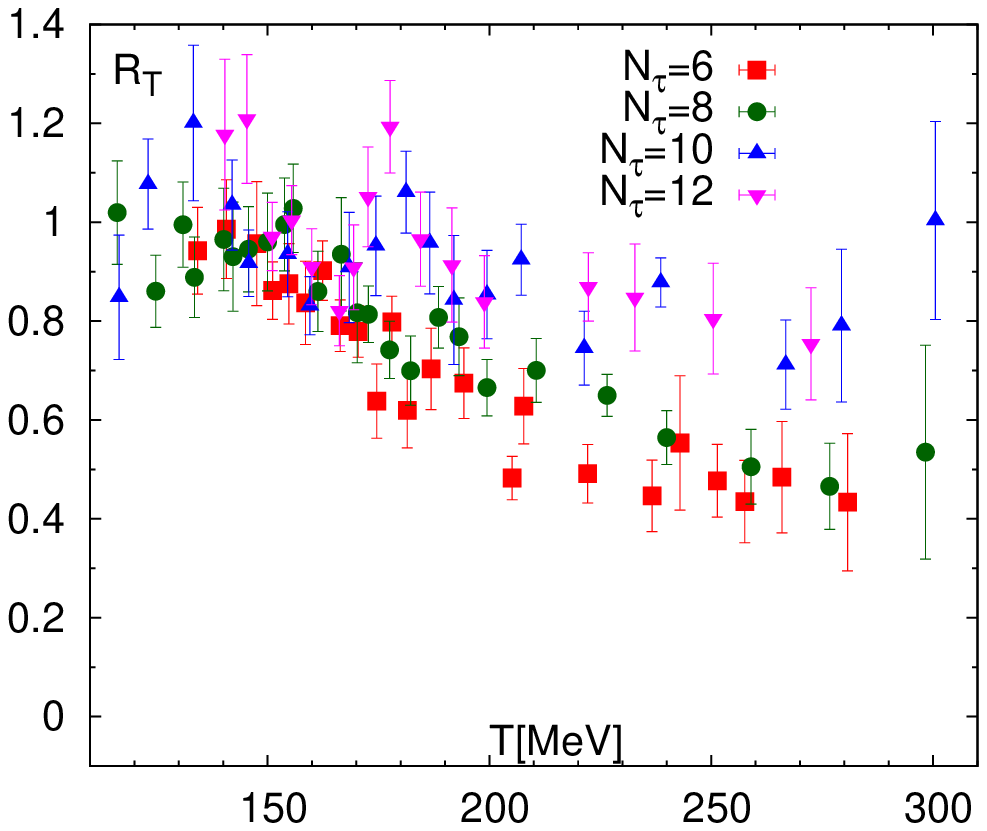}
\includegraphics[width=5.8cm]{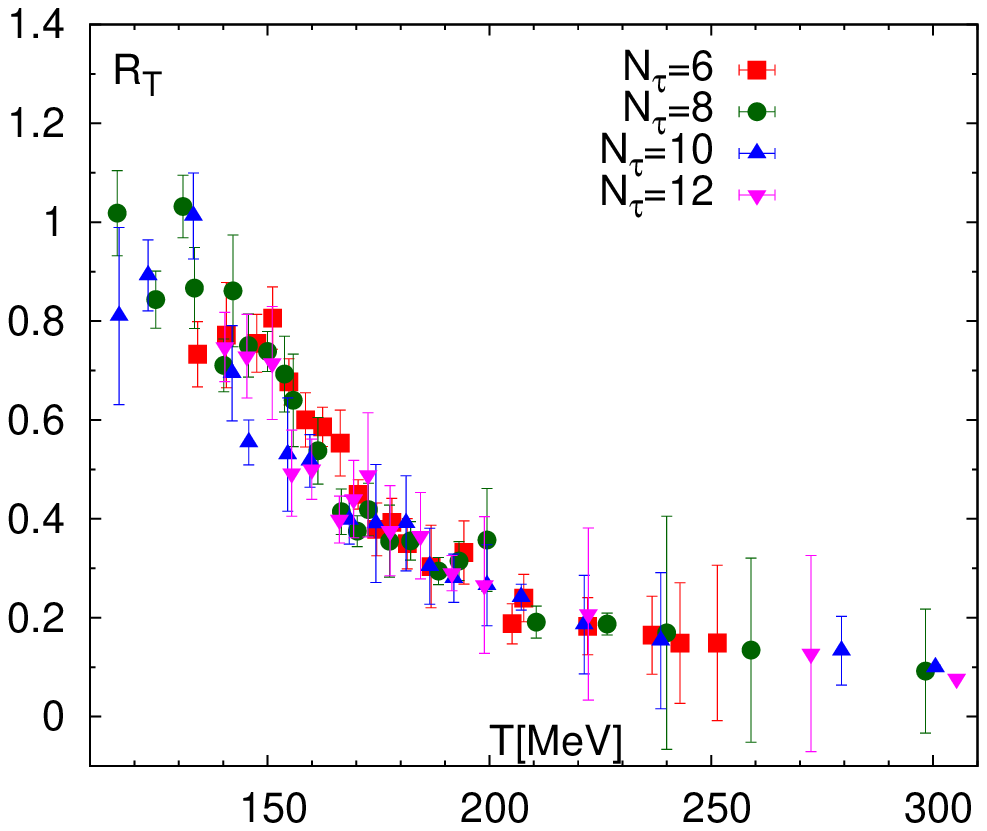}
\includegraphics[width=5.8cm]{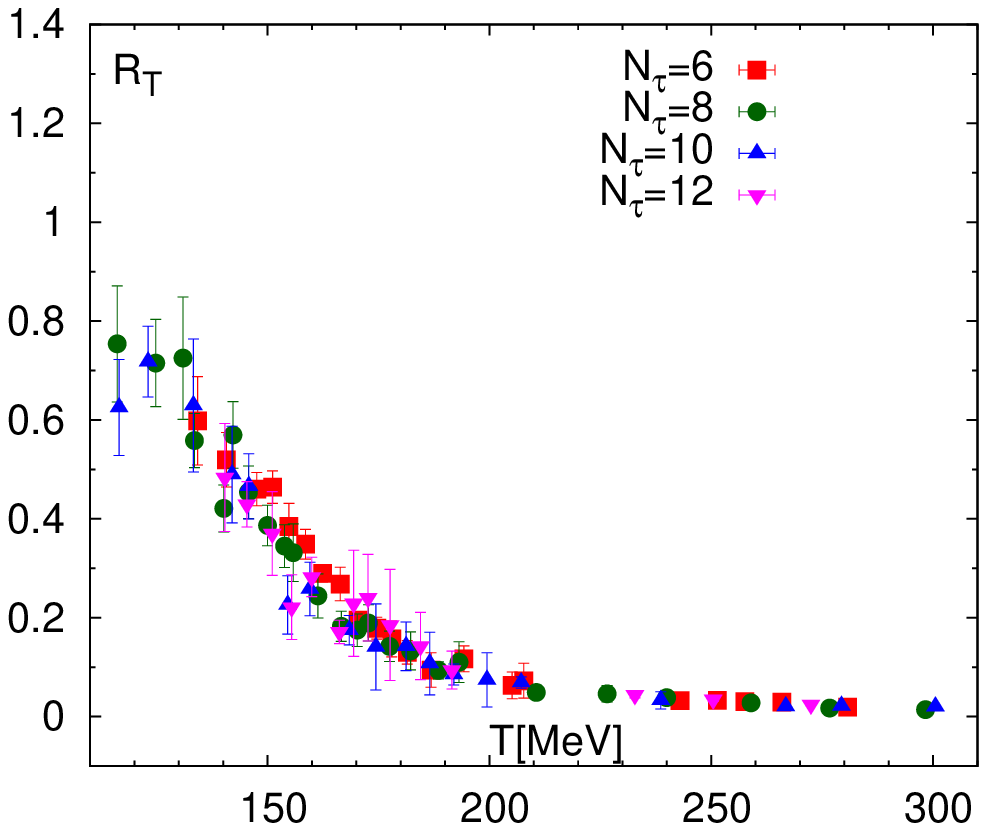}
\caption{\label{fig:ratT} 
The ratio of the susceptibilities $R_T$ shown as function of the temperature 
for zero flow time (left), flow time $f=f_0$ (middle) and for flow time 
$f=3 f_0$ (right).
}
\end{figure*}

\subsection{Fluctuations of Polyakov loop}
The Polyakov loop susceptibility defined as
\begin{equation}
\chi=(V T^3) \left(\langle |P|^2 \rangle - \langle |P| \rangle^2 \right),
\label{susc}
\end{equation}
is often used to study the deconfinement transition in SU(N) gauge theories
and for the determination of the transition temperature. 
It has a sharp peak at the pseudocritical temperature. 
It is not clear, however, how to renormalize this quantity. 
Attempts to renormalize it using the square of the renormalization factor 
of the Polyakov loop have been proposed \cite{Lo:2013etb,Lo:2013hla}.
However, apart from being ad-hoc this procedure does not remove all the UV 
divergences in the susceptibility as can be seen from the comparison of 
lattice data obtained for different $N_{\tau}$ \cite{Lo:2013hla}.
In Ref. \cite{Datta:2015bzm} the gradient flow was used in the calculation 
of the Polyakov loop susceptibilities in SU(3) gauge theory. 
The gradient flow effectively renormalizes the susceptibility and thus no 
cutoff dependence can be seen \cite{Datta:2015bzm}, but the value of the 
Polyakov susceptibility depends on the choice of the flow time.
The peak position is, however, independent of the flow time and is equal 
to the phase transition temperature \cite{Datta:2015bzm}.

We also used gradient flow to study the Polyakov loop susceptibility in 2+1 
flavor QCD.
Our results for flow time $f=3 f_0$ and different $N_{\tau}$ are shown in 
Fig. \ref{fig:chi} (left panel).
The Polyakov loop susceptibility obtained for $f=3 f_0$  shows a peak around 
$T \simeq 200$ MeV, i.e. at significantly higher temperature than  the peak position in $S_Q$, $T_S$
(\mbox{e.g.}~\mbox{$T_S(N_\tau=12)=157(6)\,{\rm MeV}$}).
The $N_{\tau}$ dependence of the Polyakov loop susceptibility is rather mild and does not show a clear tendency. 
Next, we examine the dependence of the Polyakov loop susceptibility on the 
flow time. 
In Fig. \ref{fig:chi} (right panel) we also show the flow time dependence of 
$\chi$ for $N_{\tau}=12$, where the flow time dependence is expected 
to be the mildest. 
We see that the Polyakov loop susceptibility strongly depends on the choice 
of the flow time. 
The peak position shifts to large values as the flow time is decreased 
from $3 f_0$ to $f_0$. 
This behavior of the Polyakov loop susceptibility in 2+1 flavor QCD can be 
understood as follows.
Unlike in SU(N) gauge theory the Polyakov loop is not related to singular 
behavior of the free energy in the transition region. 
The fluctuations of the Polyakov loop are therefore not affected by the 
critical behavior in the transition region and thus are not enhanced in 
a significant way. 
The value of $\chi$ is determined by the regular terms and thus depends 
on the renormalization procedure, i.e., the choice of the flow time. 

In addition to the Polyakov loop susceptibility defined by Eq. (\ref{susc}), 
which corresponds to the fluctuation in the absolute value of the Polyakov 
loop, one can consider separately the fluctuations of real and imaginary 
parts of the Polyakov loop
\begin{equation}
\chi_L=(VT)^3 \langle ({\rm Re} P)^2 \rangle-\langle P\rangle^2,~~\chi_T 
= (VT)^3 \langle ({\rm Im} P)^2 \rangle,
\end{equation}
which, following Refs. \cite{Lo:2013etb, Lo:2013hla}, we will call the 
longitudinal and transverse susceptibilities. 
In the above equations we used the fact that 
$\langle P \rangle=\langle {\rm Re} P \rangle$ and 
$\langle {\rm Im} P \rangle=0$.
We have calculated $\chi_L$ and $\chi_T$  using the gradient flow. 
We find that $\chi_L$ behaves as $\chi$, i.e. it has the same flow time 
dependence, and for $f=3 f_0$ it shows a broad peak in the temperature 
region $T=(180-200)$ MeV. 
We also find a significant flow time dependence for $\chi_T$. 
However, $\chi_T$ has a peak at temperatures around $160$ MeV, i.e. close 
to the chiral transition temperature. One may speculate that with 
increasing the flow time further the peak position of $\chi_L$ will move 
closer to the chiral transition temperature because the large flow time 
will enhance the infrared fluctuations in the real part of the Polyakov 
loop. 
However, we did not pursue this in the present study.

In Refs. \cite{Lo:2013etb,Lo:2013hla} the ratios of the Polyakov loop 
susceptibilities
$R_A=\chi/\chi_L$ and $R_T=\chi_T/\chi_L$ have been studied. 
It has been argued there that these ratios are sensitive probes of 
deconfinement and are independent of the cutoff.
Therefore, we will study these ratios in more detail. 
First, let us consider the ratio $R_A$.
It is shown in Fig. \ref{fig:ratA} as function of the temperature for 
various flow times and lattice spacings. 
For zero flow time our results for $R_A$ are in qualitative agreement 
with the results of Ref. \cite{Lo:2013hla}. 
The ratio $R_A$ exhibits a crossover behavior for temperatures 
$T=(150-200)$ MeV. However, we see a very strong cutoff ($N_{\tau}$) 
dependence of this ratio. 
While for $N_{\tau}=8$ the crossover happens at temperatures close to 
the chiral transition
temperatures, for larger $N_{\tau}$ it happens at significantly higher 
temperatures. For flow time $f=f_0$ we do not see any significant 
cutoff dependence in $R_A$, i.e. this value of the flow time is 
sufficiently large to get rid of the cutoff effects and obtain a 
renormalized quantity for $R_A$ (cf. the middle panel of 
Fig. \ref{fig:ratA}). 
Since cutoff effects are quite small already for $f=f_0$ it is 
sufficient to study the flow time dependence of our results for the 
$N_{\tau}=8$ lattice data, which is also shown in Fig. \ref{fig:ratA}. 
One can see from the figure that as the flow time increases the value 
of $R_A$ at low temperatures increases, and the step function like 
behavior of $R_A$ gradually disappears. 
For flow time $f=2 f_0$ and $f=3 f_0$ the ratio $R_A$ smoothly 
approaches one from below as the temperature increases and shows no 
sign of an inflection point. 
Note, that there is no significant flow time dependence for 
$f \ge 2 f_0$ in $R_A$.
The flow time dependence for other $N_{\tau}$ is similar.

Now let us examine the temperature dependence of $R_T$. 
In Fig. \ref{fig:ratT} we show our results for $R_T$ for three 
different flow times: $f=0, f_0$ and $3 f_0$. 
For zero flow time we see sizable cutoff dependence in $R_T$ and 
our results are qualitatively similar to those of Ref. \cite{Lo:2013hla}. 
For flow time $f=f_0$ the large cutoff dependence is removed and we 
see a crossover like behavior around temperatures of about $160$ MeV. 
For $f=3 f_0$ we have a very similar picture and again we see a crossover 
behavior around temperatures of about $160$ MeV. 
However, the value of $R_T$ is somewhat reduced at low temperatures. 

In summary, we find that the ratios $R_A$ and $R_T$ are strongly cutoff 
dependent contrary to the conjecture of Refs. \cite{Lo:2013etb, Lo:2013hla} 
stating their cutoff independence. 
Evaluating these ratios with the gradient flow removes the cutoff dependence. 
However, $R_A$ obtained from the gradient flow is not sensitive to 
deconfinement. 
On the other hand $R_T$ obtained from the gradient flow is sensitive to 
deconfinement, it shows a crossover behavior close to the chiral crossover 
temperature.
Furthermore, $R_T$ is not very sensitive to the choice of the flow time, 
and therefore it can be considered as a sensitive probe of deconfinement.
}

\section{\label{sec:weak} Comparison with the weak-coupling calculations}
{
\begin{figure}
\includegraphics[width=8cm]{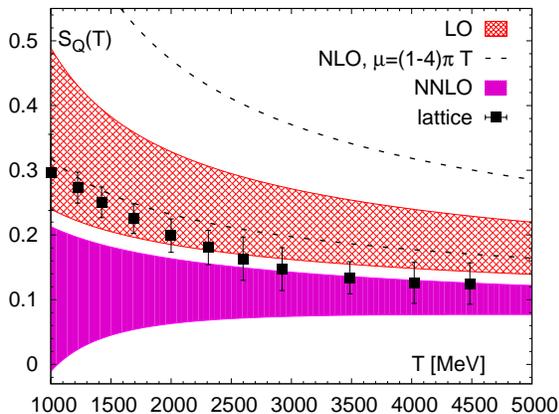}
\caption{\label{fig:weak} 
The lattice results obtained with
$N_{\tau}=4$ compared to the weak-coupling calculations.}
\end{figure}

In this section we discuss the comparison of our lattice results with 
the weak-coupling calculations. 
The free energy of a static quark has been calculated to next-to-next-to 
leading order (NNLO) \cite{Berwein:2015ayt}.
It is important to calculate the free energy to this order to reduce the 
large scale dependence of the weak-coupling result. 
We will use the $N_{\tau}=4$ results for this comparison as these extend 
up to the rather high temperatures of $5814$ MeV and the lattice artifacts 
are small, see discussions in Sec. \ref{sec:high}. 
As was pointed out in Ref. \cite{Berwein:2015ayt} the comparison of the 
lattice results
and the weak-coupling calculations is complicated by the fact that the two 
calculations are performed in different schemes. 
The weak-coupling calculations are performed in $\overline{MS}$ scheme, 
while in the lattice calculations the scheme is fixed by the prescribed 
values of the static $Q \bar Q$ energy at zero temperature at some distance. 
The two schemes can be related by a constant (temperature independent) 
shift in $F_Q$ that can be calculated.
This, however, introduces additional uncertainty in the comparison. 
The most straightforward way to perform the comparison of the lattice 
and the weak-coupling results is to consider the 
entropy \cite{Berwein:2015ayt}. 
Such a comparison has been performed in SU(3) gauge theory, i.e. for 
$N_f=0$ in a temperature range extending up to $24T_d$, with 
$T_d \simeq 300$ MeV being the deconfinement phase transition temperature 
in \cite{Berwein:2015ayt}. 
It was found that the lattice data are in between the leading order (LO) 
and the NNLO results, and at the highest temperature the NNLO and the 
lattice results agree within the uncertainties.

In Fig. \ref{fig:weak} we show the comparison of the LO and NNLO 
weak-coupling results with the $N_{\tau}=4$ results for $S_Q$. 
We used the 1-loop running coupling
constant in the weak-coupling calculations and the value 
$\Lambda_{\overline{MS}}=315$ MeV obtained from the static energy 
at zero temperature \cite{Bazavov:2014soa}. 
This value is compatible with the earlier determination from the 
static energy in Ref. \cite{Bazavov:2012ka}. 
The bands shown in Fig. \ref{fig:weak} correspond to scale variations 
between $\mu=\pi T$ and $\mu=4 \pi T$. 
At the highest temperature the lattice results and the NNLO results 
agree within the estimated uncertainties. 
At lower temperatures, $T<1500$ MeV the lattice results are closer to 
the LO weak-coupling results. 
For $T<1000$ MeV the NNLO result for $S_Q$ can turn negative for some 
choices of the renormalization scale. 
This is clearly an unphysical behavior indicating that higher-order 
corrections are too large. 
The situation is quite different from the case of quark number 
susceptibilities, where the weak-coupling prediction seems to work 
for $T>300$ MeV \cite{Bazavov:2013uja,Ding:2015fca}. 
This is due to the fact that quark number susceptibilities are dominated 
by the contribution of the non-static Matsubara modes, while for the free 
energy of a static quark the dominant contribution comes from the static 
sector \cite{Berwein:2015ayt}.
Overall, the agreement of the weak-coupling and the lattice results for 
$S_Q$ is similar to the case of the SU(3) gauge theory. 
As previously noted in Sec.~\ref{sec:high} the value of $S_Q$ at high 
temperature in QCD is larger than in the SU(3) gauge theory. 
This increase is well explained by the weak-coupling calculations. 
}

\section{Conclusions}
{
In summary, we have calculated the free energy of a static quark in 2+1 
flavor QCD at physical quark masses using several lattice spacings and 
in a large temperature range. 
We have presented continuum results for this quantity at much higher 
temperature than previously available. 
We also calculated the entropy of a static quark and showed that it is 
a useful quantity for studying deconfinement in 2+1 flavor QCD. 
Namely, we showed that it has a peak at a temperature around the chiral 
transition temperature, indicating that deconfinement and chiral 
transitions happen at similar temperatures. 
The entropy of a static quark is also useful for comparing lattice and 
weak-coupling results at high temperatures. 
Since the cutoff effects are very small at high temperatures we could 
do this comparison
using the $N_{\tau}=4$ lattice results which extend up to temperatures 
as high as
$5814$ MeV. 
At the highest temperatures we see agreement between the lattice and 
the NNLO weak-coupling results within the estimated uncertainties but at 
lower temperatures higher-order corrections become large and the 
weak-coupling expansion may not be reliable.

We also studied the fluctuations of the Polyakov loop using the gradient 
flow. 
We showed that Polyakov loop susceptibilities can be renormalized using 
the gradient flow
and the transverse Polyakov loop susceptibility may be a sensitive probe 
of deconfinement.
}

\section*{Acknowledgments}
This work was supported by U.S. Department of Energy under 
Contract No. DE-SC0012704.
We acknowledge the support by the DFG Cluster of Excellence 
``Origin and Structure of the Universe'' (Universe cluster). 
The calculations have been carried out on Blue Gene/L computer of 
New York Center for computational Science in BNL,
at NERSC, on the computing facilities of the Computational Center for 
Particle and Astrophysics (C2PAP) and
on SuperMUC cluster of the Leibniz Supercomputer Center (LRZ). 
Usage of C2PAP and SuperMUC took place under the three Universe cluster 
grants ``Static Quark Correlators in lattice QCD at nonzero temperature'' 
for 2014, 2015 and 2016 (project ID pr83pu) and the LRZ grant 
``Properties of QCD at finite temperature'' for 2015 (project ID pr48le).
N.~Brambilla, A.~Vairo and J. H.~Weber acknowledge the support by the 
Universe cluster for the seed project ``Simulating the Hot Universe'', 
by the Bundesministerium f\"{u}r Bildung und Forschung (BMBF) under grant 
``Verbundprojekt 05P2015 - ALICE at High Rate (BMBF-FSP 202) GEM-TPC Upgrade 
and Field theory based investigations of ALICE physics'' under grant 
No. 05P15WOCA1 and by the
Kompetenznetzwerk f\"{u}r Wissenschaftliches H\"{o}chstleistungsrechnen 
in Bayern (KONWIHR) for the Multicore-Software-Initiative with the project 
``Production of gauge configurations at zero and nonzero temperature'' 
(KONWIHR-IV). 
H.-P. Schadler was funded by the FWF DK W1203 
``Hadrons in Vacuum, Nuclei and Stars''.

\appendix
\section{DETAILS OF THE LATTICE CALCULATIONS}
\label{appendix:lattice_details}
{
In this appendix we will discuss the gauge configurations and the 
calculation of the bare Polyakov loop used in the present analysis 
as well as some other details of the lattice calculations.
As mentioned in Sec. \ref{sec:Setup} we have used the gauge configurations 
and the bare Polyakov loop calculated by the HotQCD Collaboration in 
Refs. \cite{Bazavov:2011nk, Bazavov:2014pvz}.
The parameters corresponding to these gauge configurations, 
including the gauge coupling $\beta$, strange quark mass and the 
accumulated statistics are given in Table VI of 
Ref. \cite{Bazavov:2011nk}
and Table III of Ref. \cite{Bazavov:2014pvz}. 
The values of the bare Polyakov loops are given in Tables X, XI
and XII of Ref. \cite{Bazavov:2011nk} and Tables IX, X, XI 
and XII of Ref. \cite{Bazavov:2014pvz}.
We also used the Polyakov loop calculated on $40^3 \times 10$ 
lattices in Ref. \cite{Bazavov:2013yv}.
To extend the calculations of the Polyakov loop to significantly 
higher temperatures we performed calculations on $16^3 \times 4$ 
lattices.
The parameters of these calculations along with the expectation 
values of the bare Polyakov loop are given in Table \ref{tab:nt4}.
We also used the gauge configurations generated for the study of 
the quark number susceptibilities in 
Refs. \cite{Bazavov:2013uja, Ding:2015fca}. 
The bare lattice parameters and the statistics corresponding to 
these gauge configurations are given in Table \ref{tab:high_beta}. 
We extended the beta range for $N_{\tau}=6,~8$ and these additional 
ensembles are also shown in Table \ref{tab:high_beta}.
The expectation values of the bare Polyakov loop are also shown 
in this Table. 
Finally, we have found it necessary to extend some of the previous 
gauge ensembles in order to have sufficiently small error for 
$L^{\rm{bare}}$. 
These ensembles with extended statistics are given in 
Table \ref{tab:ext}.
We further added a few new gauge ensembles with low beta for 
$N_{\tau}=12$, which are also included in the same Table.
Since we found for some ensembles with relatively small ensemble 
sizes that Jackknife errors are disproportionally small compared 
to ensembles with much larger ensemble sizes, we enlarged the 
respective Jackknife errors by a factor two. 
This set of ensembles with manually enlarged errors consists of 
$\beta=6.195$, $6.245$, $6.260$, $6.285$, $6.315$, $6.341$ and $6.445$ 
for $N_{\tau}=8$ and $\beta=6.990$, $7.100$ and $7.200$ for $N_{\tau}=12$. 
The criteria for enlarging the Jackknife errors for $N_{\tau}=8$ 
respectively $12$ was statistics with less than 6000 TU 
respectively 10000 TU. 
Since we did not modify the central values, these data may have a 
particularly adverse effect for the calculation of the entropy in 
the respective temperature ranges.

In Table \ref{tab:cQ} we give the values of the renormalization 
constant $c_Q$ obtained from the static energy at zero temperature. 
The renormalization constants corresponding to the direct 
renormalization are listed in Table \ref{tab:direct_renorm}.

\begin{table}
\begin{tabular}{|c|c|c|c|c|}
\hline
$\beta$ &   $am_s$  & $N_{\tau}$ & \#TU  &$L^{\rm bare}$ \\
\hline
 6.2850 &  0.079000 &  10   &  9260  & 0.000200(15) \\
 6.3410 &  0.074000 &  10   &  39220 & 0.000256(08) \\
 6.4230 &  0.067000 &  10   &  10350 & 0.000403(12) \\
 6.4450 &  0.065200 &   8   &  19150 & 0.004353(41) \\
 6.5150 &  0.060400 &  12   &  32510 & 0.000121(13) \\
 6.6080 &  0.054200 &  12   &  19890 & 0.000198(07) \\
 6.6640 &  0.051400 &  12   &  29590 & 0.000295(07) \\
 6.7000 &  0.049600 &  12   &  17070 & 0.000369(08) \\
 6.7700 &  0.046000 &  12   &  16890 & 0.000585(11) \\
 6.8400 &  0.043000 &  12   &  18720 & 0.000930(14) \\
 6.9100 &  0.040000 &  12   &  9230  & 0.001382(18) \\
\hline
\end{tabular}
\caption{
List of extended and new gauge ensembles and the corresponding 
parameters.}
\label{tab:ext}
\end{table}

\begin{table*}
\begin{tabular}{|c|c|c|c|c|c|c|c|c|c|}
\hline
$\beta$ &   $am_s$  & \multicolumn{2}{c}{$N_{\tau}=12$} & 
\multicolumn{2}{c}{$N_{\tau}=10$} &  \multicolumn{2}{c}{$N_{\tau}=8$} &
\multicolumn{2}{c}{$N_{\tau}=6$} \\
        &          & \#TU  &$L^{\rm bare}$& \# TU &$L^{\rm bare}$ 
        & \#TU &$L^{\rm bare}$&  \#TU  &$L^{\rm bare}$ \\
\hline
 7.2000 & 0.029600 &  4590  & 0.004236(34)  &  4990  & 0.013329(77)  
 &  ---   & ---            &  ---   & ---        \\
 7.5000 & 0.022200 &  8990  & 0.008988(35)  &  4990  & 0.022541(89)  
 &  4990  & 0.053915(132)  &  6670  & 0.125003(202) \\
 7.6500 & 0.019200 &  6220  & 0.011604(79)  &  2990  & 0.027463(107) 
 &  2990  & 0.062378(213)  &  ---   & ---        \\
 8.0000 & 0.014000 &  6090  & 0.019224(89)  &  39270 & 0.040274(211) 
 &  21810 & 0.083107(317)  &  4200  & 0.165828(296) \\
 8.2000 & 0.011670 &  30090 & 0.024071(46)  &  27490 & 0.047833(97)  
 &  3070  & 0.093920(224)  &  10110 & 0.181014(126) \\
 8.4000 & 0.009750 &  29190 & 0.029292(53)  &  8530  & 0.055774(114) 
 &  2990  & 0.105302(286)  &  10160 & 0.195736(153) \\
 8.5700 & 0.008376 &  3040  & 0.033996(136) &  2990  & 0.062941(191) 
 &  10260 & 0.115224(140)  &  10200 & 0.208437(141) \\
 8.7100 & 0.007394 &  3140  & 0.037736(117) &  --    & --            
 &  10040 & 0.122951(148)  &  10230 & 0.217793(125) \\
 8.8500 & 0.006528 &  2990  & 0.041652(169) &  --    & --            
 &  10010 & 0.130870(126)  &  10070 & 0.227509(159) \\
 9.060  & 0.004834 &  --    & --            &  ---   & ---           
 & 10820 & 0.142401(130)  & 10080 & 0.241385(124) \\
 9.230  & 0.004148 &  --    & --            &  ---   & ---           
 & 10260 & 0.152031(119)  & 10070 & 0.252699(154) \\
 9.360  & 0.003691 &  --    & --            &  ---   & ---           
 &  8130 & 0.158728(172)  &  8250 & 0.260614(197) \\
 9.490  & 0.003285 &  --    & --            &  ---   & ---           
 &  8020 & 0.165972(115)  &  8140 & 0.268749(158) \\
 9.670  & 0.002798 &  --    & --            &  ---   & ---           
 &  8060 & 0.174995(146)  & 10300 & 0.279328(173) \\
\hline
\end{tabular}
\caption{
The parameters and the expectation values of the bare Polyakov loops for 
the high temperature runs for $N_{\tau}=6,~8,~10$ and $12$.}
\label{tab:high_beta}
\end{table*}

\begin{table*}[t]
\parbox{.40\linewidth}{
  \begin{tabular}{|c|c|c|c|c|c|}
    \hline
    $ \beta $  & $am_s$ & T\,[\rm{MeV}] & \#TU & $ L^{\mathrm{bare}} $  \\
    \hline
    5.900 & 0.132000 &  201 & 65350 & 0.05906(11) \\
    6.000 & 0.113800 &  221 & 62610 & 0.07475(15) \\
    6.050 & 0.106400 &  232 & 62400 & 0.08280(13) \\
    6.125 & 0.096600 &  249 & 63510 & 0.09502(17) \\
    6.215 & 0.086200 &  272 & 26650 & 0.10985(21) \\
    6.285 & 0.079000 &  291 & 25380 & 0.12144(16) \\
    6.354 & 0.072800 &  311 & 19480 & 0.13242(15) \\
    6.423 & 0.067000 &  333 & 21930 & 0.14331(17) \\
    6.515 & 0.060300 &  364 & 31330 & 0.15767(14) \\
    6.575 & 0.056400 &  386 & 22770 & 0.16674(12) \\
    6.608 & 0.054200 &  399 & 39400 & 0.17192(12) \\
    6.664 & 0.051400 &  421 & 75770 & 0.17996(09) \\
    6.800 & 0.044800 &  480 & 37860 & 0.19956(10) \\
    6.950 & 0.038600 &  554 & 38090 & 0.21979(11) \\
    7.150 & 0.032000 &  669 & 31800 & 0.24452(12) \\
    7.280 & 0.028400 &  753 & 42810 & 0.25964(10) \\
    7.373 & 0.025000 &  819 & 65010 & 0.26990(10) \\
    7.500 & 0.022200 &  918 & 42950 & 0.28351(10) \\
    7.596 & 0.020200 & 1000 & 69920 & 0.29314(10) \\
    7.825 & 0.016400 & 1222 & 65380 & 0.31535(08) \\
    8.000 & 0.014000 & 1422 & 27510 & 0.33112(11) \\
    8.200 & 0.011670 & 1687 & 20790 & 0.34806(13) \\
    8.400 & 0.009750 & 1999 & 20950 & 0.36397(18) \\
    8.570 & 0.008376 & 2308 & 20280 & 0.37672(08) \\
    8.710 & 0.007394 & 2597 & 20200 & 0.38712(10) \\
    8.850 & 0.006528 & 2921 & 19210 & 0.39680(11) \\
    9.060 & 0.004834 & 3487 & 20950 & 0.41113(07) \\
    9.230 & 0.004148 & 4021 & 21240 & 0.42222(08) \\
    9.360 & 0.003691 & 4484 & 10620 & 0.43025(14) \\
    9.490 & 0.003285 & 5000 & 10320 & 0.43755(15) \\
    9.670 & 0.002798 & 5814 & 10340 & 0.44799(15) \\
    \hline
  \end{tabular}
  \caption{\label{tab:nt4}
  The parameters of $N_{\tau}=4$ ensembles and the corresponding 
  expectation values of the bare Polyakov loops.
  }
}
\hfill
\parbox{.50\linewidth}{
\begin{tabular}{|c|c|c|c|c|c|}
\hline
$\beta$ & $c_Q$ & $\beta$ & $c_Q$ & $\beta$ & $c_Q$  \\
\hline
 5.9000 & -0.3773(39) & 6.4230 & -0.4160(41) & 6.8800 & -0.3973(29) \\ 
 6.0000 & -0.3914(66) & 6.4600 & -0.4122(55) & 6.9500 & -0.3943(28) \\ 
 6.0500 & -0.3983(64) & 6.4880 & -0.4124(44) & 7.0300 & -0.3890(29) \\ 
 6.1000 & -0.3993(58) & 6.5500 & -0.4102(41) & 7.1500 & -0.3824(30) \\ 
 6.1950 & -0.4092(60) & 6.6080 & -0.4105(39) & 7.2800 & -0.3747(23) \\ 
 6.2850 & -0.4103(65) & 6.6640 & -0.4072(39) & 7.3730 & -0.3696(19) \\ 
 6.3410 & -0.4152(24) & 6.7400 & -0.4049(17) & 7.5960 & -0.3555(32) \\ 
 6.3540 & -0.4183(81) & 6.8000 & -0.4019(32) & 7.8250 & -0.3401(26) \\ 
\hline
\end{tabular}
\caption{\label{tab:cQ}
The renormalization constant $c_Q$ is obtained from the static energy 
at zero temperature.
}
\vspace{1ex}
  \begin{tabular}{|c|c|c|c|c|c|}
    \hline
    $ \beta $  & $ c_Q  $ & $ \beta $  & $ c_Q  $ 
    & $ \beta $  & $ c_Q  $  \\
    \hline
    5.9000 &      -0.3788(25)  &    6.4450 & { -0.4139(10)} 
    &    7.1500 &      -0.3830(22)  \\
    5.9500 & { -0.3853(23)} &    6.4600 &      -0.4138(14)  
    &    7.2000 & { -0.3795(08)} \\
    6.0000 &      -0.3917(18)  &    6.4880 &      -0.4135(11)  
    &    7.2800 &      -0.3754(20)  \\
    6.0250 & { -0.3942(16)} &    6.5150 & { -0.4125(19)} 
    &    7.3730 &      -0.3702(23)  \\
    6.0500 &      -0.3973(20)  &    6.5500 &      -0.4120(11)  
    &    7.5000 & { -0.3623(23)} \\
    6.0750 & { -0.3994(16)} &    6.5750 & { -0.4112(13)} 
    &    7.5960 &      -0.3560(20)  \\
    6.1000 &      -0.4012(29)  &    6.6080 &      -0.4103(20)  
    &    7.6500 & { -0.3523(14)} \\
    6.1250 & { -0.4041(22)} &    6.6640 &      -0.4081(17)  
    &    7.8250 &      -0.3403(21)  \\
    6.1500 & { -0.4052(15)} &    6.7000 & { -0.4067(10)} 
    &    8.0000 & { -0.3297(27)} \\
    6.1750 & { -0.4069(18)} &    6.7400 &      -0.4048(08)  
    &    8.2000 & { -0.3179(23)} \\
    6.1950 &      -0.4084(22)  &    6.7700 & { -0.4034(08)} 
    &    8.4000 & { -0.3062(21)} \\
    6.2150 & { -0.4091(16)} &    6.8000 &      -0.4019(15)  
    &    8.5700 & { -0.2965(22)} \\
    6.2450 & { -0.4111(16)} &    6.8400 & { -0.3996(11)} 
    &    8.7100 & { -0.2894(21)} \\
    6.2600 & { -0.4116(16)} &    6.8800 &      -0.3976(07)  
    &    8.8500 & { -0.2825(24)} \\ 
    6.2850 &      -0.4120(20)  &    6.9100 & { -0.3961(06)} 
    &    9.0600 & { -0.2708(24)} \\
    6.3150 & { -0.4134(13)} &    6.9500 &      -0.3940(16)  
    &    9.2300 & { -0.2618(24)} \\
    6.3540 &      -0.4135(20)  &    6.9900 & { -0.3913(07)} 
    &    9.3600 & { -0.2562(22)} \\
    6.3900 & { -0.4140(18)} &    7.0300 &      -0.3892(08)  
    &    9.4900 & { -0.2504(22)} \\
    6.4230 &      -0.4136(24)  &    7.1000 & { -0.3854(07)} 
    &    9.6700 & { -0.2431(22)} \\
    \hline
  \end{tabular}
  \caption{\label{tab:direct_renorm}
  The renormalization constant $c_Q$ from the direct 
  renormalization procedure. 
  }
}
\end{table*}

We used the gradient flow to calculate the renormalized Polyakov loop 
expectation value and the Polyakov loop susceptibilities. 
We always used step size $dt=0.01$ in lattice units in our gradient 
flow study. 
The parameters of the gradient flow analysis, including the values of 
$\beta$, the number of gauge configurations analyzed and the maximal 
flow time $t_{\rm max}$ are given in Tables \ref{tab:flow_nt6}, 
\ref{tab:flow_nt8}, \ref{tab:flow_nt10} and \ref{tab:flow_nt12}.

\begin{table*}
\parbox{.45\linewidth}{
\begin{tabular}{|c|c|c|}
		\hline
		$\beta$ & $t_{\rm{max}}$ & \# TU \\
		\hline
		5.850 & 0.850 & 5000 \\ 
		5.900 & 0.900 & 5000 \\ 
		5.950 & 1.000 & 5000 \\ 
		6.000 & 1.100 & 5000 \\ 
		6.025 & 1.150 & 5000 \\ 
		6.050 & 1.200 & 5000 \\ 
		6.075 & 1.250 & 5000 \\ 
		6.100 & 1.300 & 5000 \\ 
		6.125 & 1.350 & 5000 \\ 
		6.150 & 1.450 & 5000 \\ 
		6.175 & 1.500 & 5000 \\ 
		6.195 & 1.550 & 5000 \\ 
		6.215 & 1.700 & 5000 \\ 
		6.245 & 1.700 & 5000 \\ 
		6.285 & 1.850 & 5000 \\ 
		6.341 & 2.050 & 5000 \\ 
		\hline
	\end{tabular}
\begin{tabular}{|c|c|c|}
		\hline
		$\beta$ & $t_{\rm{max}}$ & \# TU \\
		\hline
		6.354 & 2.150 & 5000 \\ 
		6.423 & 2.400 & 5000 \\ 
		6.488 & 2.750 & 5000 \\ 
		6.515 & 2.900 & 5000 \\ 
		6.550 & 3.100 & 5000 \\ 
		6.575 & 3.250 & 5000 \\ 
		6.608 & 3.450 & 5000 \\ 
		6.664 & 3.800 & 5000 \\ 
		6.800 & 4.900 & 5000 \\ 
		6.950 & 6.500 & 5000 \\ 
		7.150 & 1.100 & 5000 \\ 
		7.280 & 1.400 & 1000 \\ 
		7.373 & 1.650 & 1000 \\ 
		7.500 & 2.050 & 1000 \\ 
		7.596 & 2.400 & 1000 \\ 
		7.825 & 3.550 & 1000 \\
		\hline
	\end{tabular}
	\caption{\label{tab:flow_nt6}
 $24^3 \times 6$ gauge configurations used for the gradient flow analysis.
 }
 }
 \hfill
\parbox{.45\linewidth}{
\begin{tabular}{|c|c|c|}
    \hline
		$\beta$ & $t_{\rm{max}}$ & \# TU \\
		\hline
		6.050 & 1.200 & 5000 \\ 
		6.125 & 1.350 & 5000 \\ 
		6.175 & 1.500 & 5000 \\ 
		6.195 & 1.550 & 5000 \\ 
		6.245 & 1.700 & 5000 \\ 
		6.260 & 1.750 & 5000 \\ 
		6.285 & 1.850 & 5000 \\ 
		6.315 & 1.950 & 5000 \\ 
		6.341 & 2.050 & 5000 \\ 
		6.354 & 2.100 & 5000 \\ 
		6.390 & 2.250 & 5000 \\ 
		6.423 & 2.400 & 5000 \\ 
		6.445 & 2.500 & 5000 \\ 
		6.460 & 2.550 & 5000 \\ 
		6.488 & 2.700 & 5000 \\ 
		6.515 & 2.850 & 5000 \\
		6.550 & 3.050 & 5000 \\ 
		\hline
	\end{tabular}
\begin{tabular}{|c|c|c|}
    \hline
		$\beta$ & $t_{\rm{max}}$ & \# TU \\
		\hline
		6.575 & 3.200 & 5000 \\ 
		6.608 & 3.400 & 5000 \\ 
		6.664 & 3.800 & 5000 \\ 
		6.740 & 4.400 & 5000 \\ 
		6.800 & 4.900 & 5000 \\ 
		6.880 & 5.700 & 5000 \\ 
		6.950 & 6.550 & 5000 \\ 
		7.030 & 7.550 & 5000 \\ 
		7.150 & 5.000 & 1000 \\ 
		7.280 & 5.000 & 1000 \\ 
		7.373 & 5.000 & 1000 \\ 
		7.500 & 2.100 & 1000 \\ 
		7.596 & 5.000 & 1000 \\ 
		7.825 & 5.000 & 1000 \\ 
		8.000 & 25.000 & 1000 \\ 
		8.200 & 6.700 & 1000 \\ 
		8.400 & 9.400 & 1000 \\
		\hline
	\end{tabular}
	\caption{\label{tab:flow_nt8}
$32^3 \times 8$ gauge configurations used for the gradient flow analysis.
}
}
\end{table*}

\begin{table*}
\parbox{.45\linewidth}{
\begin{tabular}{|c|c|c|}
    \hline
		$\beta$ & $t_{\rm{max}}$ & \# TU \\
		\hline
		6.285 & 1.850 & 2420 \\ 
		6.341 & 2.050 & 5000 \\ 
		6.423 & 2.450 & 3640 \\ 
		6.488 & 2.700 & 5000 \\ 
		6.515 & 2.850 & 5000 \\ 
		6.575 & 3.200 & 5000 \\ 
		6.608 & 3.400 & 5000 \\ 
		6.664 & 3.800 & 5000 \\ 
		6.700 & 4.050 & 4000 \\ 
		6.740 & 4.400 & 5000 \\ 
		6.770 & 4.650 & 4460 \\ 
		6.800 & 4.900 & 5000 \\ 
		6.840 & 5.300 & 4580 \\ 
		6.880 & 5.700 & 9720 \\ 
		\hline
	\end{tabular}
\begin{tabular}{|c|c|c|}
    \hline
		$\beta$ & $t_{\rm{max}}$ & \# TU \\
		\hline
		6.950 & 6.500 & 5000 \\ 
		7.030 & 7.550 & 5000 \\ 
		7.150 & 9.450 & 5000 \\ 
		7.200 & 11.500 & 1000 \\ 
		7.280 & 12.000 & 1000 \\ 
		7.373 & 1.600 & 1000 \\ 
		7.500 & 2.100 & 4000 \\ 
		7.596 & 2.400 & 1000 \\ 
		7.650 & 2.650 & 1000 \\ 
		7.825 & 3.500 & 1000 \\ 
		8.000 & 4.800 & 1000 \\ 
		8.200 & 6.700 & 1000 \\ 
		8.400 & 9.400 & 1000 \\ 
		8.570 & 12.500 & 1000 \\
		\hline
	\end{tabular}
  \caption{\label{tab:flow_nt10}
$40^3 \times 10$ gauge configurations used for the flow analysis.
}
}
\hfill
\parbox{.45\linewidth}{
\begin{tabular}{|c|c|c|}
    \hline
		$\beta$ & $t_{\rm{max}}$ & \# TU \\
		\hline
		6.664 & 3.800 & 3230 \\ 
		6.700 & 4.100 & 5000 \\ 
		6.740 & 4.400 & 5000 \\ 
		6.770 & 4.650 & 5000 \\ 
		6.800 & 4.900 & 5000 \\ 
		6.840 & 5.300 & 5000 \\ 
		6.860 & 5.500 & 5000 \\ 
		6.880 & 5.700 & 5000 \\ 
		6.910 & 6.050 & 4120 \\ 
		6.950 & 6.500 & 5000 \\ 
		6.990 & 7.050 & 5000 \\ 
		7.030 & 7.550 & 5000 \\ 
		7.150 & 9.450 & 1000 \\ 
		\hline
	\end{tabular}
\begin{tabular}{|c|c|c|}
    \hline
		$\beta$ & $t_{\rm{max}}$ & \# TU \\
		\hline
		7.200 & 10.350 & 3600 \\ 
		7.280 & 11.950 & 1000 \\ 
		7.373 & 14.150 & 5000 \\ 
		7.500 & 2.050 & 1000 \\ 
		7.596 & 2.400 & 5000 \\ 
		7.650 & 2.650 & 1000 \\ 
		7.825 & 5.000 & 1000 \\ 
		8.000 & 5.000 & 1000 \\ 
		8.200 & 7.000 & 1000 \\ 
		8.400 & 10.000 & 1000 \\ 
		8.570 & 12.500 & 1000 \\ 
		8.710 & 15.800 & 1000 \\ 
		8.850 & 19.950 & 1000 \\
		\hline
	\end{tabular}
  \caption{\label{tab:flow_nt12}
$48^3 \times 12$ gauge configurations used for the gradient flow
analysis.
}
}
\end{table*}
}

\section{INTERPOLATIONS AND EXTRAPOLATIONS}
\label{appendix:fits}
{
\begin{figure*}[bht]
\includegraphics[width=8cm]{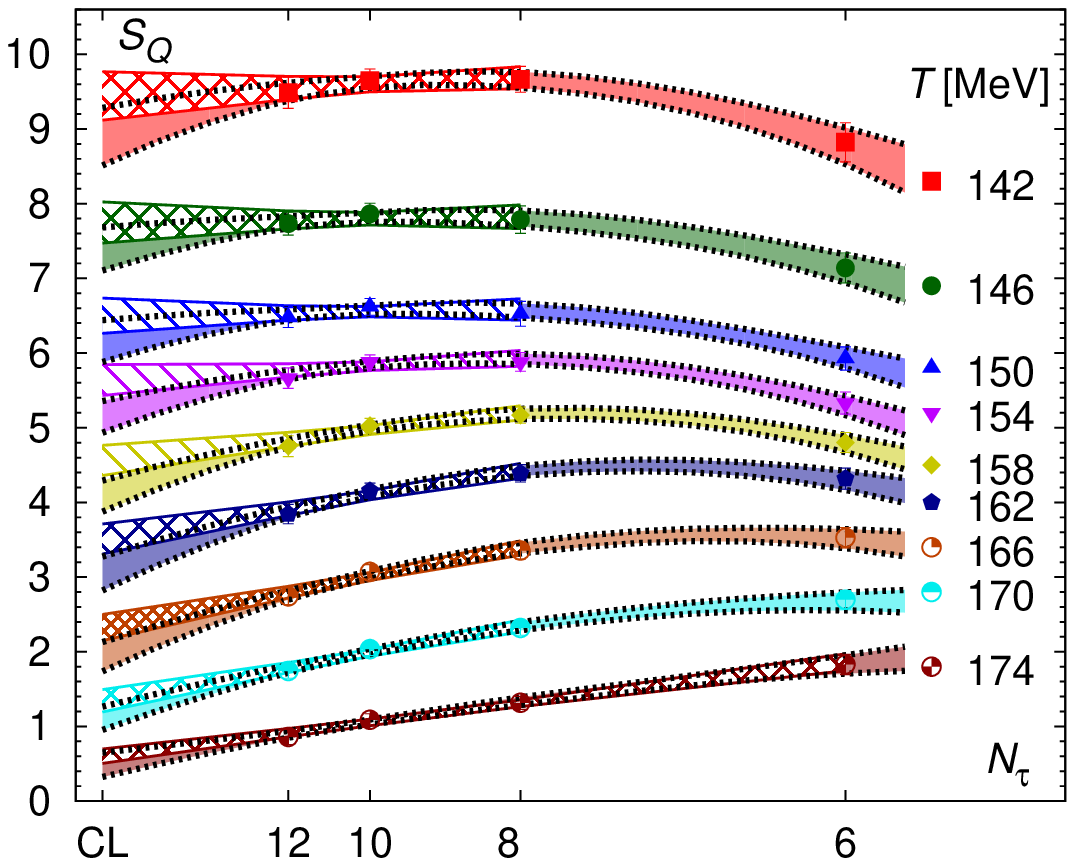}
\includegraphics[width=8cm]{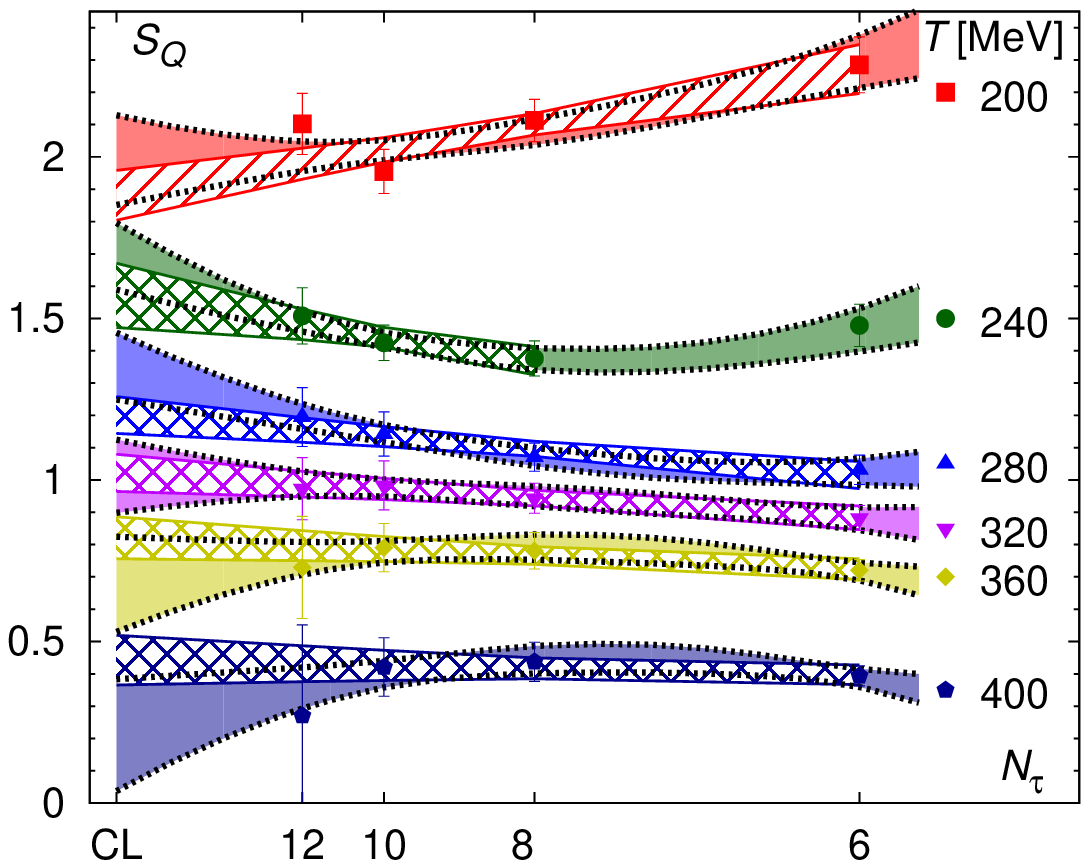}
\caption{\label{fig:SNt} 
The static quark entropy at various temperatures as function of 
$1/N_{\tau}^2$. 
``CL'' marks the continuum limit ($N_\tau \to \infty$). 
The $1/N_{\tau}^2$ continuum extrapolations are shown as bands with 
filled pattern. 
The continuum extrapolations with $1/N_{\tau}^4$ term included are shown 
as solid filled bands.
The widths of the band shows the statistical uncertainty of the fits.
The left panel shows the results in the low temperature region, while the 
right panel shows the result in the high temperature region.}
\end{figure*}

In this appendix we present some details of our interpolation procedure. 
As discussed in the main text we use polynomial fits and smoothing splines 
for the interpolations. 
The calculations of $c_Q$ and the corresponding interpolations are 
performed in three steps. 
In the first step we interpolate the value of $c_Q$ obtained in the 
$Q\bar Q$ procedure in the interval $\beta=5.900-7.825$. 
Then we use $c_Q$ obtained in the direct renormalization procedure and 
interpolate in the interval $\beta=5.900-8.850$. 
Finally we calculate $c_Q$ at higher $\beta$ using direct renormalization 
only and interpolate in the interval $\beta=5.900-9.67$.
The details of the interpolations are given in Table \ref{tab:inter_cQ}. 
In the Table, $ n_k $ is the number of knots for spline interpolations and 
${\rm sm}$ is the smoothing parameter for the built-in smooth spline 
interpolations of the R statistical package~\cite{Rpackage}.
$ n_p $ is the polynomial order for polynomial interpolations.
We refer to the interpolation of $c_Q$ from $Q\bar Q$ procedure as 
the 0th iteration of direct renormalization.

To calculate the entropy we also perform interpolations of the bare free 
energy in $\beta$. 
The details of these interpolations are presented in 
Table \ref{tab:inter_fQbare}. 
The column labels are the same as in Table \ref{tab:inter_cQ}.

In some temperature ranges, continuum extrapolations do not converge well 
and yield $\chi^2/{\rm df}>1$. 
First, in the temperature interval $176$ MeV $<T<189$ MeV, local 
continuum extrapolations of $F_Q$ with $N_{\tau}\geq 8$ yield up to 
$\chi^2/{\rm df}=1.23$ (\mbox{cf.} \mbox{Fig.}~\ref{fig:fNt}). 
Second, in the temperature interval $150$ MeV $<T<169$ MeV, local 
continuum extrapolations of $S_Q$ yield up to $\chi^2/{\rm df}=1.50$ 
with $N_{\tau}\geq8$ and $P_4= 0$.
Third, in the temperature interval $190$ MeV $<T<211$ MeV, local 
continuum extrapolations of $S_Q$ yield up to $\chi^2/{\rm df}=3.37$ 
with $N_{\tau}\geq6$ and $P_4\neq 0$ and up to $\chi^2/{\rm df}=3.69$ 
with $N_{\tau}\geq8$ and $P_4= 0$.
Judging from \mbox{Fig.}~\ref{fig:SNt}, these poor continuum extrapolations 
are caused by fluctuations of some $N_{\tau}=12$ data in the interval 
$190$ MeV $<T<211$ MeV, which originate in the relatively small ensemble 
sizes underlying some data in this interval (\mbox{cf.} 
Appendix~\ref{appendix:lattice_details}). 

We summarize the global fits in Table~\ref{tab:global fits}. Hereby, 
$n_{i},\ i=0,2,4$ are the orders of the temperature polynomials $P_i(T)$ 
as in \mbox{Eq.}~(\ref{eq:global fit}). 
We include in the table ratios 
$R[N_{\tau}]=\chi^2[N_{\tau}]/n_{\rm pt}[N_{\tau}]$ 
as measure how well data for each $N_{\tau}$ is matched by the global fit. 
Global residuals of $\chi^2/{\rm df} \lesssim 0.3$ are required to bring 
all ratios $R[N_{\tau}]$ sufficiently below one such that global fits 
yield reliable results. 

We collect the final, continuum extrapolated results for the free energy 
and entropy in Table \ref{FQ and SQ}.

\begin{table}[b]
   \begin{tabular}{|c|c|c|c|c|c|}
     \hline
     Scheme & $\beta$ & $ n_k $, $\mathrm sm$ & $\frac{\chi^2}{\mathrm df}$ 
     & $n_p$ & $\frac{\chi^2}{\mathrm df}$ \\
     \hline
     \multicolumn{6}{|l|}{$Q\bar Q$ procedure} \\
     \hline
      0th iteration & [5.900,7.825] & 5, 0.18  & 0.30 &  &  \\
     \hline
     \multicolumn{6}{|l|}{Direct renormalization} \\
     \hline
      1st iteration & [5.900,8.850] & 5, 0.04  & 0.83 &  &  \\
      2nd iteration & [5.900,9.670] & 6, 0.03  & 0.28 & 5 & 0.17 \\
     \hline
   \end{tabular}
   \caption{Spline and polynomial interpolations of $ c_Q $.
    }
   \label{tab:inter_cQ}
\end{table}

\begin{table*}[hbt]
\parbox{.49\linewidth}{
   \begin{tabular}{|c|c|c|c|c|c|c|}
    \hline
    $N_{\tau}$ & $T$ [MeV] & $\beta$ & $ n_k $, $\mathrm sm$ 
    & $\frac{\chi^2}{\mathrm df}$ & $n_p$ & $\frac{\chi^2}{\mathrm df}$ \\ 
    \hline
     4 & [201,5814] & [5.900,9.670] & 18, 0.0  & 0.93 & 9 & 0.71 \\
    \hline
     6 & [234,237]  & [5.900,6.488] &  7, 0.05 & 0.90 & 9 & 0.79 \\
     6 & [181,3876] & [6.215,9.670] & 19, 0.0  & 1.03 & 9 & 1.06 \\
    \hline
     8 & [116,227]  & [6.050,6.740] &  6, 0.0  & 1.01 & 7 & 0.73 \\
     8 & [201,2907] & [6.515,9.670] & 15, 0.0  & 0.87 & 9 & 1.07 \\
    \hline
    10 & [116,239]  & [6.285,7.030] &  6, 0.0  & 0.86 & 6 & 0.83 \\
    10 & [181,924]  & [6.740,8.570] &  8, 0.0  & 1.06 & 7 & 1.01 \\
    \hline
    12 & [122,233]  & [6.515,7.200] &  5, 0.05 & 0.88 & 4 & 0.85 \\
    12 & [185,974]  & [6.950,8.850] &  8, 0.0  & 0.96 & 7 & 1.00 \\
    \hline
   \end{tabular}
   \caption{\label{tab:inter_fQbare}
   Primary spline and polynomial interpolations of $ f_Q^{\mathrm{bare}} $.
   }
}
\hfill
\parbox{.49\linewidth}{   
     \begin{tabular}{|c|c|c|c|c|c|c|c|}
    \hline
    $\{n_{0},n_{2},n_{4}\}$ & $ N_{\tau}^{\rm min}$ & $T$ [MeV] &
    $\frac{\chi^2}{\mathrm df}$ & $R[12]$ & $R[10]$ & $R[8]$ & $R[6]$  \\ 
    \hline
    $ \{6,5,0\} $ & 8 & [115,215] & 0.26 & 0.30 & 0.26 & 0.22 & --   \\
    $ \{6,5,4\} $ & 6 & [115,225] & 0.24 & 0.21 & 0.30 & 0.34 & 0.10 \\
    \hline
    $ \{5,3,0\} $ & 8 & [173,410] & 0.21 & 0.36 & 0.25 & 0.04 & --   \\
    $ \{5,3,0\} $ & 6 & [173,410] & 0.28 & 0.62 & 0.17 & 0.27 & 0.10 \\
    \hline
  \end{tabular}
  \caption{\label{tab:global fits}
  Global continuum extrapolations using $Q\bar Q$ procedure.
  The last four columns denote 
  $R[N_{\tau}] =\frac{\chi^2(N_{\tau})}{n_{\rm pt}(N_{\tau})}$, the ratio of 
  residues and number of points for each $N_{\tau}$.
  }
}  
\end{table*}

\begin{table*}[t]
\parbox{.49\linewidth}{
  \begin{tabular}{|c|c|c|}
    \hline
    $T$ [MeV] & $F_Q$ [MeV] & $S_Q$  
    \\
    \hline
    125  & 481(7) &    1.99(56) \\
    130  & 470(6) &    2.00(49) \\
    135  & 459(5) &    2.29(40) \\
    140  & 446(4) &    2.78(34) \\
    145  & 431(4) &    3.26(29) \\
    150  & 413(4) &    3.60(24) \\
    155  & 395(4) &    3.67(20) \\
    160  & 377(4) &    3.47(20) \\
    165  & 360(4) &    3.19(18) \\
    170  & 345(4) &    2.94(15) \\
    175  & 330(4) &    2.83(15) \\
    180  & 317(4) &    2.75(15) \\
    185  & 303(4) &    2.56(15) \\
    190  & 291(4) &    2.45(13) \\
    \hline
    200  & 268(4) &    2.15(13) \\
    210  & 248(4) &    1.89(11) \\
    220  & 231(4) &    1.72(12) \\
    230  & 214(5) &    1.67(10) \\
    240  & 198(5) &    1.60(10) \\
    250  & 182(5) &    1.52(10) \\
    260  & 170(4) &    1.38(08) \\
    270  & 157(5) &    1.30(08) \\
    \hline
  \end{tabular}
  \begin{tabular}{|c|c|c|}
    \hline
    $T$ [MeV] & $F_Q$ [MeV] & $S_Q$ 
    \\
    \hline
    280  & 144(5) &    1.21(07) \\
    290  & 132(5) &    1.14(06) \\
    \hline
    300  & 121(6) &    1.07(06) \\
    310  & 113(5) &    0.99(06) \\
    320  & 103(6) &    0.94(06) \\
    330  &  94(6) &    0.90(06) \\
    340  &  85(6) &    0.87(05) \\
    350  &  76(6) &    0.85(05) \\
    360  &  68(7) &    0.82(05) \\
    370  &  59(7) &    0.80(05) \\
    380  &  50(6) &    0.80(05) \\
    390  &  42(6) &    0.77(05) \\
    \hline
    400  &  34(6) &    0.75(05) \\
    410  &  27(6) &    0.74(05) \\
    420  &  19(6) &    0.72(05) \\
    430  &  12(7) &    0.71(05) \\
    440  &   6(7) &    0.69(05) \\
    450  &  -1(7) &    0.68(05) \\
    460  &  -8(7) &    0.66(05) \\
    470  & -14(7) &    0.64(05) \\
    480  & -20(8) &    0.63(05) \\
    490  & -26(8) &    0.61(05) \\
    \hline
  \end{tabular}

}
\hfill
\parbox{.49\linewidth}{
  \begin{tabular}{|c|c|c|}
    \hline
    $T$ [MeV] & $F_Q$ [MeV] & $S_Q$ 
    \\
    \hline
    500  & -32(8)   &   0.60(06)  \\
    520  & -44(8)   &   0.57(06)  \\
    540  & -55(8)   &   0.55(06)  \\
    560  & -66(9)   &   0.53(06)  \\
    580  & -76(9)   &   0.51(06)  \\
    \hline
    600  &  -86(9)  &   0.49(07)  \\
    620  &  -96(9)  &   0.47(07)  \\
    640  & -105(9)  &   0.45(06)  \\
    660  & -114(10) &   0.43(06)  \\
    680  & -122(11) &   0.41(06)  \\
    \hline
    700  & -130(12) &   0.41(04) \\
    720  & -138(12) &   0.40(05) \\
    740  & -145(13) &   0.40(05) \\
    760  & -152(13) &   0.39(05) \\
    780  & -159(14) &   0.38(05) \\
    \hline
    800  & -165(14) &   0.38(05) \\
    820  & -171(14) &   0.37(05) \\
    840  & -177(15) &   0.36(05) \\
    860  & -182(15) &   0.36(05) \\
    880  & -187(15) &   0.35(05) \\
    \hline
    900  & -192(15) &   0.34(05) \\
    920  & -196(16) &   0.33(05) \\
    \hline
  \end{tabular}
  \begin{tabular}{|c|c|c|}
    \hline
    $T$ [MeV] & $F_Q$ [MeV] & $S_Q$ 
    \\
    \hline
    940  & -202(20)    &   0.33(04) \\
    960  & -209(20)    &   0.32(04)  \\
    980  & -215(21)    &   0.32(04) \\
    \hline
    1000  & -221(21)   &   0.31(04)  \\
    1100  & -251(24)   &   0.28(03)  \\
    1200  & -277(25)   &   0.24(04)  \\
    1300  & -301(25)   &   0.23(04)  \\
    1400  & -323(25)   &   0.21(04)  \\
    1500  & -343(27)   &   0.19(04)  \\
    1600  & -362(28)   &   0.18(05)  \\
    1700  & -390(30)   &   0.17(06)  \\
    1800  & -397(31)   &   0.16(06)  \\
    1900  & -413(32)   &   0.16(06)  \\
    \hline
    2000  & -430(32) &   0.16(06) \\
    2400  & -501(34) &   0.17(03) \\
    2800  & -584(47) &   0.15(02) \\
    3200  & -664(45) &   0.14(03) \\
    3600  & -717(47) &   0.13(04) \\
    4000  & -768(49) &   0.12(03) \\
    4400  & -817(52) &   0.12(03) \\
    4800  & -867(51) &   0.13(03) \\
    5200  & -929(54) &   0.15(06) \\
    \hline
  \end{tabular}  
}
  \caption{\label{FQ and SQ}
  Continuum limit of the free energy $F_Q$ and entropy $S_Q$ for high 
  temperatures. 
  $F_Q$ is a shifted finite $N_\tau$ result above $T>920$ MeV. 
  For $T \leq 2800$ MeV, ${N_\tau}=8$ is used and for $T > 2800$ MeV 
  $N_{\tau}=4$ is used. 
  The cutoff effects at $T=920$ MeV are used as shift and added to the 
  errors linearly. 
  $S_Q$ is a finite $N_\tau$ result above $T>680$ MeV. 
  For $T \leq 2000$ MeV, ${N_\tau}=8$ is used and for $T > 2000$ MeV 
  $N_{\tau}=4$ is used. 
  Errors of $S_Q$ for these $N_{\tau}$ are increased by 
  $0.01$ to account for systematic effects.
  }
\end{table*}
}

%

\end{document}